\documentclass[aps,pre,twocolumn,showpacs,superscriptaddress]{revtex4-2}
	\usepackage[utf8]{inputenc}
	\usepackage[T1]{fontenc}
	\usepackage{lmodern}
	\usepackage{amsmath,amssymb,amsfonts}
	\usepackage{bm}
	\usepackage{graphicx}
	\usepackage{hyperref}
	\usepackage{booktabs}
	
	\usepackage{lineno}
	
\begin{document}
		
		\title{Critical slip distance on rough faults}
		
		\author{Davide Zaccagnino}
		\affiliation{Institute of Risk Analysis, Prediction and Management (Risks-X), Academy for Advanced Interdisciplinary Studies, Southern University of Science and Technology (SUSTech), 1088 Xueyuan Rd., 518055, Shenzhen, Guangdong, China}
		\affiliation{National Institute of Geophysics and Volcanology (INGV), Via di Vigna Murata 605, 00143, Roma, Italy}
		
		\author{Giacomo Pozzi}
		\affiliation{Dipartimento di Geoscienze, Università degli Studi di Padova, Via Gradenigo 6, 35131 Padova, Italy}
		
\begin{abstract}
	The critical slip distance $D_c$ - the characteristic displacement over which a fault dynamically weakens from peak to residual strength - is a key parameter in earthquake rupture dynamics, controlling the nucleation process and slip evolution. Nevertheless, its physical origin and scaling remain debated: $D_c$ is observed to span from micrometers in laboratory experiments to meters on natural faults, a scaling which cannot be explained within the standard friction framework, which interprets $D_c$ as a material constant related to the contact population and provides no mechanism linking it to fault structure or event size.
	Here, we propose a first-principles derivation showing that $D_c$ is a structural property of faults, governed by their self-affine roughness belonging to the Kardar-Parisi-Zhang universality class. 
	The framework unifies all scales through a single expression where 
	$D_c = \max(\delta_{\rm frac}, \frac{C}{2\mu} l_{\rm eff}^{\zeta})$, 
	with $l_{\rm eff}$ given by the minimum of rupture length and pulse width, $\mu$ is the static friction, while $C$ and $\zeta$ represent the roughness amplitude and exponent, respectively. 
	The intrinsic material length, $\delta_{\rm frac}$, emerges as the fracture process zone size, which is dominated by geometric unlocking at macroscopic scales. We further demonstrate that rate-and-state friction emerges as the mean-field limit of the underlying fractal asperity dynamics and provide analytical formulas for its parameters: the state variable represents the population-averaged contact age, while the critical distance identifies the geometric unlocking of the dominant pinning bumps.
	Rate-and-state friction is therefore an emergent phenomenology whose mathematical form reflects the statistical mechanics of the underlying contact population, not material rheology, and its parameter values are strictly scale-dependent.
	This allows us to better characterize fault stability from a multiscale perspective; hence, we introduce a new concept named \textquotedblleft fault retentivity'': the ability of a fault to arrest a nucleated rupture before it degenerates into run-away via the multiscale barrier population encoded in the fault structure. Retentivity determines whether a fault hosts only small earthquakes or can occasionally nucleate large ones. The key implication is that assessing fault stability cannot rest on friction alone: it requires to understand the hierarchical architecture of the fault system.
\end{abstract}
		\maketitle
	\section*{Introduction}
	
	\subsection*{The problem of upscaling $D_c$}
	Earthquake rupture dynamics is controlled by the evolution of frictional strength on the fault surface. A key quantity in this process is the critical slip distance, $D_c$, which characterizes the amount of slip that must accumulate for the shear strength on fault to drop from its static value to its residual (dynamic) one during the early stages of rupture nucleation and propagation. This slip-weakening distance determines the fracture energy, the nucleation size; likely, it affects rupture velocity and the ground motion produced by earthquakes \cite{Cocco2002,Tinti2005}.
	
	Despite its importance, $D_c$ is still a poorly understood quantity whose physical meaning is often underrated because of its pragmatic employment as one of the crucial fit parameters of the rate-and-state friction laws.  Laboratory friction experiments (centimeter to decimeter scale) measure $D_c$ values in the range of 1-100 microns  \cite{Dieterich1979,Marone1998}. Conversely, seismological inversions of natural earthquakes (kilometer to 
	hundred-kilometer scale) routinely infer $D_c$ values of centimeters to 
	meters, and for the largest events, several meters 
	\cite{Papageorgiou1983,Ellsworth2003}. 
	
This represents a scale amplification of five to six orders of magnitude 
for which no existing constitutive framework provides a complete account.
Only a part of the discrepancy may arise from the fact that laboratory and seismological studies often refer to different operative definitions of critical distance. 
Indeed, in rate-and-state friction, $D_c$ is a characteristic length calibrated from velocity-step tests where the slip speed changes by about one order of magnitude; it is proportional to the contact renewal distance but is smaller than the total slip required to reduce friction from its peak static value to the dynamic level.  
Conversely, the seismological slip-weakening distance measures the full breakdown from peak to residual strength over the much larger velocity changes that occur during earthquake rupture.  
Thus, a careful comparison that uses the same definition of $D_c$ may reduce the laboratory-field gap, but a scale dependence of several orders of magnitude persists, demanding a physical explanation that goes beyond the empirical fitting of rate-and-state parameters.
 
Early efforts to address the scaling of $D_c$ recognised that the critical slip distance on natural faults should reflect the characteristic size of the asperities that lock the interface. The first attempt we are aware dates back to \cite{Scholz1988}, exploiting the then-emerging evidence for fractal fault roughness, computed the expected size of the largest contact junctions and obtained $D_c \sim 1-10$ mm for mature faults at seismogenic depths - a prediction that, 
at the time, appeared broadly consistent with the limited seismological estimates available. As \cite{Marone1998} subsequently cautioned, models of this kind assume direct contact between bare rock surfaces and therefore neglect the presence of fault gouge, a granular wear product that is ubiquitous on mature faults and that modifies the contact population, the frictional response, and the characteristic slip distance. However, the accumulation of high-quality 
seismological and geodetic source inversions over the intervening two 
decades has completely reshaped the observational picture: $D_c$ for 
large crustal and subduction earthquakes is now routinely inferred to be in the order of meters \cite{Ide1997,Tinti2005,Galetzka2015}. These values exceed the theoretical predictions by two to three orders of magnitude. 

The difficulty of reconciling laboratory friction with field observations extends beyond the scaling of $D_c$ \cite{Zaccagnino2026} and highlights a broader tension in the current understanding of fault strength \cite{Zaccagnino2025}. Purely frictional descriptions, 
in which fault stability is assessed solely through the rate-and-state 
parameters $a$ and $b$, have proven increasingly insufficient to account for the observed spectrum of slip behaviour. Faults that are nominally frictionally stable under laboratory-derived criteria can host earthquakes, while rate-and-state formulations that successfully describe centimetre-scale experiments fail to reproduce the complexity of rupture on geometrically mature, gouge-bearing faults \cite{DalZilio2023,Barbery2025,Li2025}. Recent works have 
demonstrated that the transition from aseismic to seismic slip is 
controlled not by frictional parameters alone but by the structural 
architecture of the fault zone - including its roughness, gouge thickness, and degree of localization \cite{Ross2020,Cochran2023,Lee2024}. These findings indicate that friction, as traditionally formulated, is an incomplete description of fault behavior, and that a proper account of the geometric structure of the interface is required to understand how ruptures nucleate, propagate, and arrest \cite{Harbord2017, Dong2024}.

It is within this broader context that the scaling of $D_c$ and its physical meaning must be reconsidered.

	\subsection*{$D_c$ as a fault frictional property}
	The standard theoretical framework for fault friction is the rate-and-state 
	formalism \cite{Dieterich1979,Ruina1983,Rice1983}. In its most common form, 
	the shear stress $\tau$ on the fault is expressed as
	\begin{equation}
		\tau = \sigma_n \left[ \mu_0 + a \ln\left(\frac{V}{V_0}\right) + 
		b \ln\left(\frac{V_0 \theta}{D_c}\right) \right],
		\label{eq:RSintro}
	\end{equation}
	where $\sigma_n$ is the effective normal stress, $V$ is the sliding 
	velocity, $V_0$ is a reference velocity, $\mu_0$ is the steady-state 
	friction coefficient at velocity $V_0$, $a$ and $b$ are empirical 
	rate-and-state parameters, and $\theta$ is a state variable representing 
	the average contact age or maturity. 
	The state variable evolves according 
	to one of several empirically motivated laws, typically the aging law
	\begin{equation}
		\frac{d\theta}{dt} = 1 - \frac{V\theta}{D_c},
		\label{eq:agingIntro}
	\end{equation}
	or the slip law
	\begin{equation}
		\frac{d\theta}{dt} = -\frac{V\theta}{D_c} \ln\left(\frac{V\theta}{D_c}\right).
		\label{eq:slipLawIntro}
	\end{equation}
	In this framework, $D_c$ appears as a constitutive parameter - the slip 
	distance required to renew the contact population - which does not depend on the scale of the rupture. $D_c$ is then treated as a material property, presumably related to the characteristic asperity size or the average contact diameter. 
	Indeed, rate-and-state is a scale-free theory by construction, which cannot explain the observed several-orders-of-magnitude discrepancy between laboratory and natural fault values of $D_c$. 
	
	\subsection*{$D_c$ as a fault structural property}
	We propose a radically different perspective: $D_c$ is not an 
	intrinsic material constant, but rather a structural property of 
	the fault surface that emerges from its multiscale geometric roughness. 
	
\begin{figure}
		\includegraphics[width=\columnwidth]{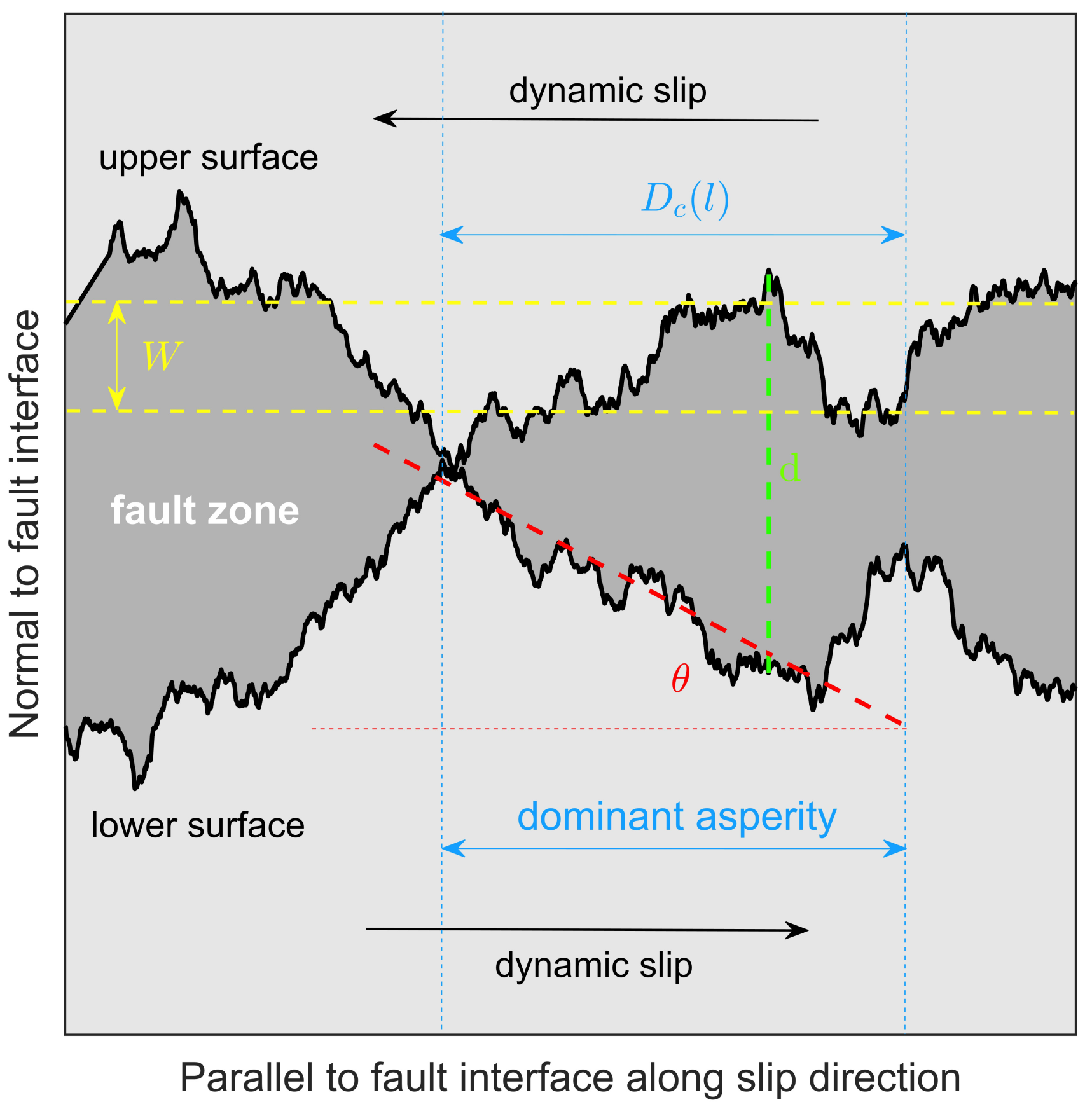}
		\caption{Schematic representation of the fault zone as the interface of two shifting pinned self-affine surfaces. The dominant (largest) asperity controls friction (via the angle $\theta$) at the scale $l$. The critical slip distance, $D_c$, is the minimum sliding length allowing the system to overcome the mechanical barrier due to fault roughness.}
		\label{fig:concept}
	\end{figure}

	The central physical picture is illustrated schematically in 
	Fig.~\ref{fig:concept}. 
	
	A fault is not a flat plane but a rough interface with topography spanning from micron-scale grain boundaries to kilometer-scale fault scarps. When two such rough volumes are pressed together under tectonic load, they interlock: peaks from one surface nest into valleys of the other. For slip to occur, these interlocked asperities must be either fractured through or geometrically unlocked. This perspective has been already successfully employed in theoretical studies and simulations \cite{DeRubeis1996, Milanese2019}. 
	The critical slip distance is precisely the horizontal 
	displacement required to unlock the largest bump that participates in the rupture process.
	This geometric interpretation naturally implies why $D_c$ scales with event size: small laboratory samples contain only small bumps (limited by the sample dimensions), so the unlocking distance is small. 
	Large earthquakes rupture patches that contain much larger bumps (up to the seismogenic depth), requiring correspondingly larger slip to unlock. 
	The scale dependence is a direct consequence of the self-affine, fractal nature of fault roughness.
	
	\subsection*{Toward a new view of fault stability}
	The theory we develop is based on three physical ingredients, which together provide a possible explanation for the observed trend of the critical slip distance at different scales.
	
	The first is the structure of the fault surface. Faults are not flat planes but self-affine interfaces \cite{Schmittbuhl1993}, rough on all scales from micrometres, where microstructures control frictional properties \cite{Pozzi2022}, to kilometres \cite{Fang2013}. This geometry obeys well-defined scaling laws that fix its statistical properties and can be captured by a single parameter, i.e., the roughness amplitude. The population of asperities that lock the interface - their number, steepness, and strength - follows directly from this description, providing the structural foundation for all that follows.
	
	The second is the mechanics of asperity failure. For a fault to slip, the opposing surfaces must overcome the barriers that pin them together. At the smallest scales, these barriers are cohesive \cite{Weiss2016}: mineral cements and cold-welded contacts that must be fractured. At larger scales, they are geometric \cite{Gabrielov1996, Nielsen1998}: interlocked bumps that must be climbed. The transition between these two regimes introduces a characteristic length below which the critical slip distance is a constant, material-controlled quantity, and above which it becomes scale-dependent, growing with the size of the bumps involved. For the largest earthquakes, this growth saturates, capped by the finite thickness of the seismogenic layer.
	
	The third is the hierarchical nature of rupture arrest. A growing rupture does not interact with a single barrier but a statistical sequence of them, distributed over all scales according to the fault roughness. Whether the rupture arrests or cascades to a larger event is governed by the extreme value statistics of this population. This cascade connects the smallest scales, where fracture dominates, to the largest, where geometric unlocking controls the displacement, and it provides the bridge between laboratory friction and earthquake dynamics.
	
	Taken together, these three ingredients allow us to build a unified expression for $D_c$ that spans the full range of observations. 
	
	They also reveal that standard rate-and-state friction is nor a fundamental neither a constitutive law, but the emergent ensemble average of the underlying fractal contact population. From this perspective, the parameters that govern fault stability are not empirical constants to be fitted, but measurable properties of the fault surface itself.
	
	\section*{Faults as self-affine interfaces}\label{sec:roughness}
	A self-affine surface is one whose statistical properties are invariant 
	under an anisotropic scaling transformation: if we rescale the horizontal 
	coordinates by a factor $\lambda$, we must rescale the vertical height 
	by a different factor $\lambda^H$ to obtain a statistically 
	identical surface. The exponent $H$ is the Hurst roughness exponent, which characterizes the degree of correlation in its structure.
	High-resolution topographic measurements of exhumed fault surfaces, 
	obtained from laboratory analyses (micron to millimeter scale), terrestrial LiDAR (millimeter to meter scale), and satellite or drone-based photogrammetry (meter to kilometer scale), reveal how faults show self-affine appearance over an extraordinarily wide range of scales 
	\cite{Power1987,Schmittbuhl1995,Sagy2007,Candela2009,Brodsky2011}.
	 
	Then, we can apply the mathematical formalism of self-affine objects to faults in order to get information about their physical behavior. 
	
	Let $h(\mathbf{x})$ be the height of the fault surface at the 
	two-dimensional position $\mathbf{x} = (x, y)$, where $x$ is the direction 
	of slip and $y$ is the perpendicular (fault-normal) coordinate. We focus 
	on one-dimensional profiles along the slip direction, denoted $h(x)$ with 
	$y$ held fixed. For a stationary self-affine random field, the fundamental 
	statistical descriptor is the structure function
	\begin{equation}
		S(L) \equiv \left\langle \left[ h(x+L) - h(x) \right]^2 \right\rangle,
		\label{eq:structureFunc}
	\end{equation}
	where $\langle \cdot \rangle$ denotes the spatial average along the profile. The root-mean-square height fluctuation over a horizontal lag $L$ is then
	\begin{equation}
		w(L) \equiv \sqrt{S(L)} = \left\langle \left[ h(x+L) - h(x) \right]^2 
		\right\rangle^{1/2}; 
		\label{eq:width}
	\end{equation}
	this quantity is usually called \textquotedblleft roughness''. 
	For a self-affine interface, the structure function obeys a power law
	\begin{equation}
       w(L) = C L^\zeta,
	\label{eq:powerLaw}
	\end{equation}
	where $C$ is the roughness amplitude. 
	The power-law scaling in Eq.~(\ref{eq:powerLaw}) implies that the surface has no characteristic horizontal scale: bumps of all sizes exist, with their typical height growing as a power of their base length.
	
	\subsection*{The Hurst exponent and the KPZ universality class}
	Several studies have found that exhumed fault surfaces, 
	particularly those that have accommodated significant cumulative slip, exhibit a Hurst (or roughening) exponent in the range $\zeta \approx 0.5$--$0.8$, with a central value compatible with $\zeta = 2/3 \approx 0.67$ \cite{Schmittbuhl1995,Sagy2007,Candela2009}. This value carries an important physical meaning: it places fault roughness in the Kardar-Parisi-Zhang (KPZ) universality class of growing interfaces \cite{KPZ1986}.
	
\begin{figure*}
	\includegraphics[width=2\columnwidth]{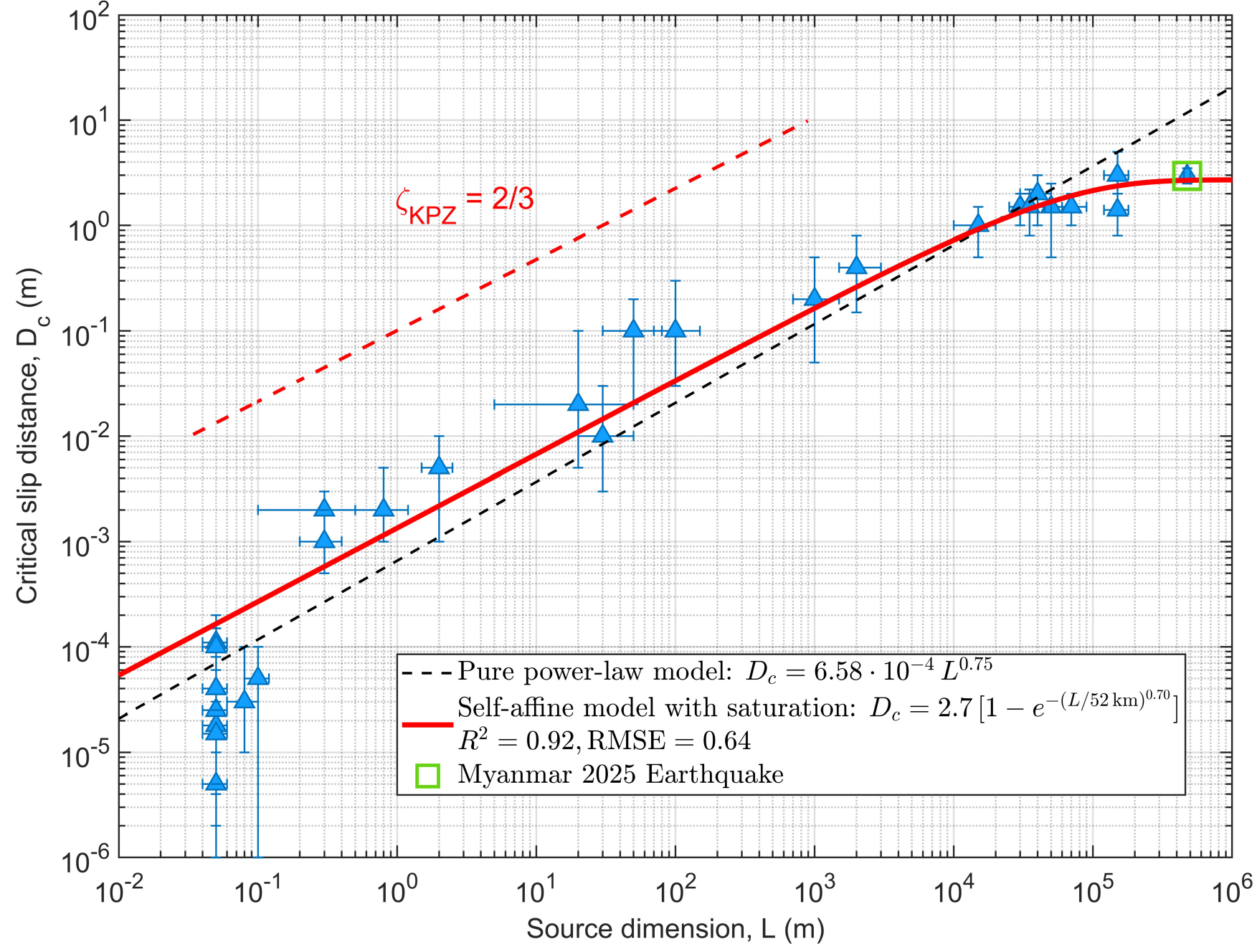}
	\caption{Scaling behavior of the critical slip distance, $D_c$ as a function of the rupture length of the seismic source. The trend is well represented by a power law with exponent $\zeta \sim 0.7$ with saturation for large events in the form of a stretched exponential (saturation length $L_0 \approx 30-40$ km). The roughness exponent is compatible with the Kardar-Parisi-Zhang theoretical prediction $\zeta_{\text{KPZ}} = 2/3$. The values of $D_c$ in the lab show systematic underestimation due to the different operative definition from stick-slip procedures (read the introduction about this topic). Data from \cite{Dieterich1979,Tullis1986,Kilgore1993,Gibowicz1994,Marone1998,
			Ide1997,Mair1999,Ohnaka1999,Ohnaka2000,Richardson2002,Mikumo2003,
			Tinti2005,Chambon2006,Galetzka2015,Leeman2016,Scuderi2017,Rubino2017,
			Ando2018,Ikari2019,McLaskey2019,Latour2025}}
	\label{fig:scaling}
    \end{figure*}

	The KPZ equation describes the stochastic growth of an interface driven by a combination of random deposition, surface tension, and nonlinear lateral growth; it reads as  
	\begin{equation}
		\frac{\partial h}{\partial t} = \nu \nabla^2 h + \frac{\lambda}{2} 
		(\nabla h)^2 + \eta(\mathbf{x}, t),
		\label{eq:KPZ}
	\end{equation}
	where $\nu$ is the surface tension, $\lambda$ is the nonlinear coupling coefficient, and $\eta$ is a Gaussian white noise with 
	$\langle \eta(\mathbf{x},t) \eta(\mathbf{x}',t') \rangle = 2D 
	\delta(\mathbf{x} - \mathbf{x}') \delta(t-t')$. The KPZ equation predicts that in one dimension (i.e., for a 1D profile of the 2D surface), the roughness exponent is exactly $\zeta = 2/3$ \cite{KPZ1986,Barabasi1995}.
	The physical interpretation is the following: faults evolve through the cumulative effect of slip events, each of which removes material from some asperities and deposits it elsewhere (through wear, gouge formation). Over geological timescales, this process drives the fault structure toward the KPZ attractor, explaining the observed $\zeta \approx	2/3$. See Figure~\ref{fig:scaling}.  
	
	Therefore, in this article we adopt $\zeta = 2/3$ as the theoretically justified value supported by observations (the empirical value of the roughening exponent retrieved from published studies is indeed compatible with $\zeta = 2/3$). 
	
	Differently, the roughness amplitude $C$ is a scale-invariant property of a 
	given fault: it does not depend on the spatial scale length $L$, but is a 
	fixed constant for that fault segment, determined by the rock type, the 
	cumulative slip history, and the wear processes that have shaped the 
	surface. Its physical meaning can be easily understood: if we set $L = 1$ m in Eq.~(\ref{eq:powerLaw}), then $w(1~\mathrm{m}) = C$. Hence, $C$ represents the root-mean-square height fluctuation over a 1-meter horizontal distance.
	Measured values of $C$ for large natural faults fall in the range $C \approx 10^{-3} \text{ to } 10^{-2} \; \mathrm{m}^{1/3}$. 
	A fault with $C = 3 \times 10^{-3}~\mathrm{m}^{1/3}$ has an RMS height 
	difference of $3$ mm over 1 m or, equivalent to and 3 m over 10 km. 
	Mature, smooth faults such as the San Andreas, the North Anatolian or the Sagaing Faults, tend toward the lower end of this range; rougher, less evolved faults occupy the upper end \cite{Brodsky2011}.
	
	An equivalent description of the roughness is through the Fourier power 
	spectral density $P(k)$ of the height profile. 
	For a self-affine surface 
	with Hurst exponent $\zeta$
	\begin{equation}
		P(k) \equiv \langle |\tilde{h}(k)|^2 \rangle \propto k^{-1 - 2\zeta},
		\label{eq:PSD}
	\end{equation}
	where $\tilde{h}(k) = \int h(x) e^{-ikx} dx$ is the Fourier transform, 
	and $k$ is the wavenumber (inverse length scale). For $\zeta = 2/3$, we have 
	$P(k) \propto k^{-7/3}$. The roughness amplitude $C$ determines the 
	prefactor of this power law. The relation between $C$ and the spectral 
	amplitude is then given by
	\begin{equation}
		C = \sqrt{\frac{1}{2\pi} \int_{2\pi/L_{\max}}^{2\pi/l_{\min}} P(k) \, dk},
		\label{eq:CfromPSD}
	\end{equation}
	with appropriate cutoffs at the minimum and maximum roughness scales.
	
	\subsection*{Anisotropy of fault roughness}
	The self-affine structure of fault surfaces is known to be anisotropic: the roughness exponent measured parallel to the slip direction often differs from that measured perpendicular to it. High-resolution measurements on exhumed faults \citep{Sagy2007, Candela2009, Brodsky2011} have shown that the slip-parallel Hurst exponent $\zeta_{\parallel}$ is smaller than the slip-perpendicular one $\zeta_{\perp}$, with typical values $\zeta_{\parallel} \approx 0.5$--$0.7$ and $\zeta_{\perp} \approx 0.6$--$0.8$. This anisotropy arises from the directionality of wear processes. Indeed, sliding along the fault smooths asperities mostly in the slip direction, while slip-perpendicular irregularities persist more.  
	
	In the present work, we focus on the slip-parallel roughness, as it is the geometry that controls the dilatant (geometric) unlocking of asperities during fault slip. The slip-perpendicular roughness, while important for the three-dimensional contact area and fluid flow, unlikely contribute to the critical slip distance $D_c$ scaling, which is measured along the direction of motion. 
	
	Therefore, the scaling exponent $\zeta$ used throughout this paper refers to the slip-parallel Hurst exponent (compatible with observations reported in the literature), and we adopt the KPZ value $\zeta = 2/3$ as theoretically justified above.
	
	\section*{The critical slip distance as the measure of geometric unlocking}
	\label{sec:geometric}
	Consider two identical rough surfaces described by Eq.~(\ref{eq:powerLaw}) 
	placed in contact with opposite orientations, so that the peaks of one surface perfectly nest into the 
	valleys of the other. This represents the fully locked fault state during the interseismic phase. 
	
	The surfaces are subjected to a normal stress $\sigma_n$ (effective normal stress, after subtracting pore pressure), and the macroscopic shear resistance is due to the combination of flat-contact friction and the dilatant work required to lift the surfaces over the interlocked bumps.
	
	Consider a single characteristic bump of base length $l$ and height 
	$w = w(l) = C l^\zeta$. Because the two surfaces are identical and perfectly 
	interlocked, the bump from the upper surface nests exactly into a 
	corresponding depression of the lower surface. 
	To nucleate slip, the system must do dilatant work against the normal load: the surfaces must be lifted apart by the full height $w$ as the bump climbs out of its socket.
	We model the bump geometry as a triangular sawtooth (a reasonable first-order approximation that captures the essential physics while permitting a simple analytical treatment). The sawtooth has base length $l$, height $w$, and a constant slope angle $\theta$ given by
	\begin{equation}
		\tan\theta = \frac{w}{l} = \frac{C l^\zeta}{l} = C l^{\zeta-1}.
		\label{eq:slope}
	\end{equation}
	For $\zeta = 2/3$, we have $\zeta-1 = -1/3$, so $\tan\theta = C l^{-1/3}$. 
	Smaller bumps are steeper; larger bumps are shallower: a direct 
	consequence of self-affinity in agreement with observations; indeed, large mature faults appear globally smoother than smaller rough faults.  
	
	When the upper surface slides horizontally by a distance $s$, the 
	vertical separation between the surfaces (the dilation) increases. At 
	slip $s$, the peak of the upper bump has ridden partway up the flank 
	of the lower bump. The shear stress required to sustain this slip is 
	the sum of the flat friction (which acts on the real area of contact) 
	and the dilatant term, given by
	\begin{equation}
		\tau(s) \approx \mu \sigma_n + \sigma_n \tan\alpha(s),
		\label{eq:tauS}
	\end{equation}
	where $\alpha(s)$ is the instantaneous dilatation angle at slip $s$, 
	which depends on the bump shape. For the sawtooth, $\alpha(s) = \theta$ 
	(approximately constant during the climbing phase).
	
    The critical state occurs when the bump has been fully unlocked: the 
	peak of the upper bump has reached the valley of the lower bump. 
	In the simplest, naive view, the slip required to achieve this is half the base length of the bump, because the system must traverse from a peak-on-peak to a peak-on-valley configuration. However, this wrong estimate 
	($D_c \approx l/2$) would lead to $D_c \propto L$, which predicts 
	kilometer-scale $D_c$ for kilometer-scale faults, which would be inconsistent with the observed few meters.
	This early estimate is incorrect because the steepest bumps on a fault surface are not stable under tectonic loading. If a bump is too steep (slope $\theta$ exceeding the macroscopic friction angle $\phi = \arctan\mu$), it will fail spontaneously: the shear stress needed to climb it exceeds the frictional strength of a flat surface, and the bump will be sheared off during the interseismic period or during 
	previous earthquakes. Therefore, the surviving bump population 
	has its slope bounded by $\phi$:
	\begin{equation}
		\tan\theta \lesssim \tan\phi = \mu.
		\label{eq:slopeBound}
	\end{equation}
	For typical rocks in the lab, $\mu \approx 0.6$-$0.85$, so $\phi \approx 
	30^\circ$--$40^\circ$.
	The bump that controls the unlocking at scale $l$ is the steepest one 
	that can survive, i.e., with $\tan\theta \approx \mu$. Its height is 
	$w = C l^\zeta$, and its base length $\lambda$ (along the slip direction) 
	is related to its height by the limiting slope:
	\begin{equation}
		\mu = \tan\theta \approx \frac{w}{\lambda} = \frac{C l^\zeta}{\lambda}
		\quad \Longrightarrow \quad \lambda = \frac{C l^\zeta}{\mu}.
		\label{eq:lambdaFromSlope}
	\end{equation}
	The critical slip distance to unlock this bump (peak to valley) is half 
	the base length:
	\begin{equation}
		D_c^{\mathrm{geom}}(l) = \frac{\lambda}{2} = \frac{C}{2\mu} \, l^\zeta.
		\label{eq:DcGeom}
	\end{equation}
	For $\zeta = 2/3$, this gives:
	\begin{equation}
		D_c^{\mathrm{geom}}(l) = \frac{C}{2\mu} \, l^{2/3}.
		\label{eq:DcGeom23}
	\end{equation} 
	This formula states that the critical slip 
	distance for geometric unlocking of a bump of size $l$ is proportional 
	to $l^{2/3}$, with a prefactor combining the roughness amplitude $C$ 
	and the friction coefficient $\mu$.
	For a small bump ($l = 1$ mm, $l^{2/3} \approx 0.01$ m$^{2/3}$, 
	$D_c \approx 10$--$100~\mu$m), the required slip is microscopic, like in 
	the laboratory \cite{Dieterich1979,Tullis1986,Marone1998,Chambon2006,Kilgore1993,Mair1999,Leeman2016,Rubino2017,Scuderi2017,Ikari2019,McLaskey2019}. 
	For a large bump ($l = 1$ km, 
	$l^{2/3} \approx 100$ m$^{2/3}$, $D_c \approx 0.1$--$1$ m), 
	the required slip is macroscopic -- this is the earthquake regime, as 
	found for moderate to large seismic events \cite{Gibowicz1994, Ide1997, Ohnaka1999,Ohnaka2000,Richardson2002,Tinti2005,Mikumo2003,Galetzka2015,Ando2018,Latour2025}.
	However, the scaling exponent $2/3$ implies that $D_c$ grows sublinearly 
	with the bump size. Relative to the bump size, indeed, the critical slip 
	decreases: $D_c/l \propto l^{-1/3}$. A millimeter bump requires 
	slip comparable to its size ($D_c/l \approx 0.1$--$1$), while a 
	kilometer bump requires only $D_c/l \approx 10^{-3}$--$10^{-4}$ of its 
	size. This is because larger bumps are smoother, and 
	the slip needed to unlock them is limited by the friction angle, not 
	by the bump full width. See Figure~\ref{fig:scaling}. 
	
	\begin{figure}
		\includegraphics[width=\columnwidth]{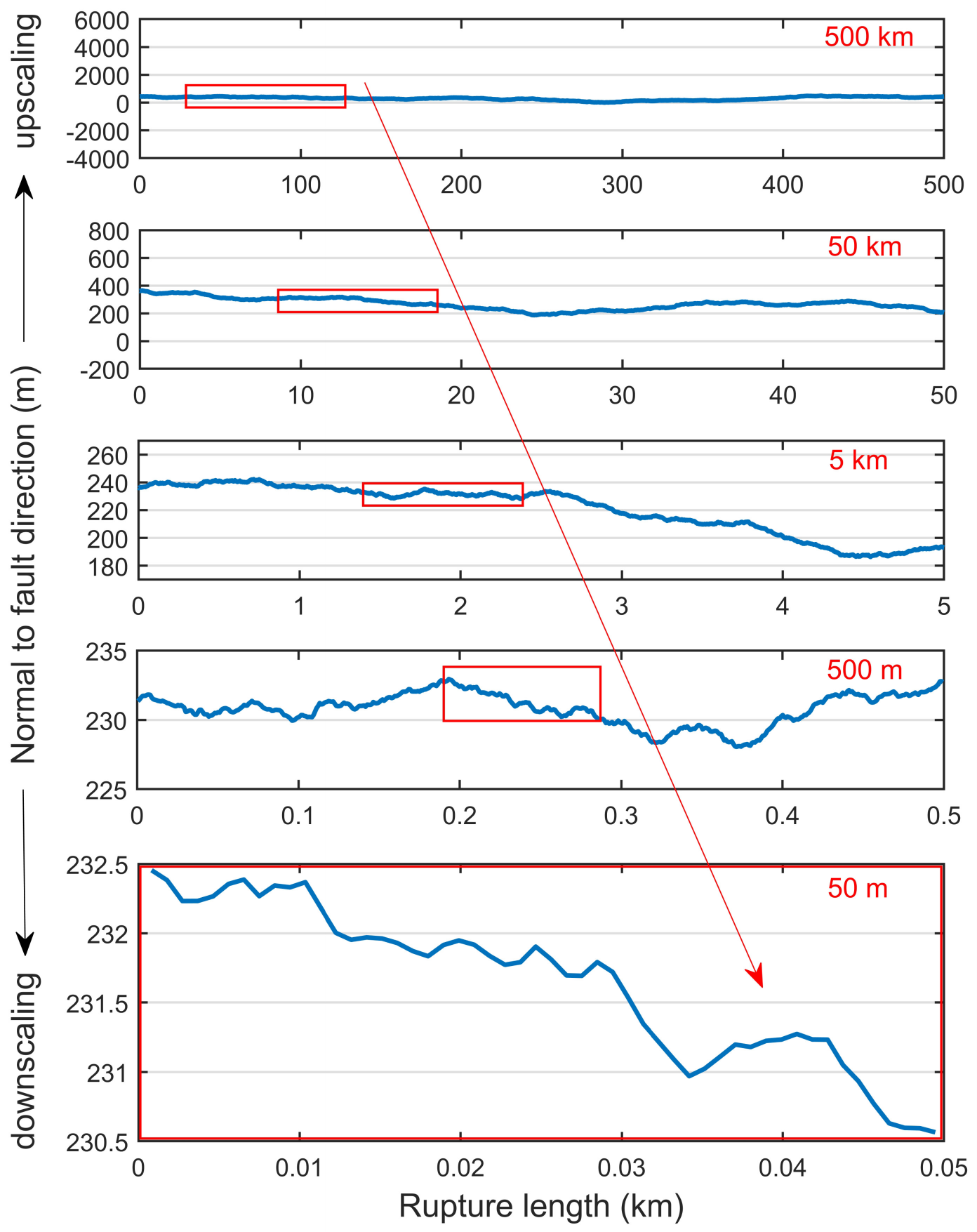}
		\caption{The effect of self-affinity on the structural properties of large and small faults: major megathrusts and elongated strike-slip faults (extension up to a few thousand kilometers) appear smooth; conversely, immature shorter faults are visually rougher.}
		\label{fig:upscaling}
	\end{figure}
	
	\section*{Saturation of the effective scale}
	\label{sec:pulse}

	The geometric derivation in the previous section expresses $D_c$ 
	as a function of the bump size $l$. But what sets $l$ for a given 
	earthquake? The critical slip distance is controlled by the largest bump 
	that the rupture must unlock. This is not necessarily the entire fault 
	length; it is the size of the region that is actively slipping 
	simultaneously during the rupture process.

We introduce the effective structural length $l_{\mathrm{eff}}$, 
defined as the linear dimension of the slipping patch whose asperities 
collectively determine the peak-to-residual strength drop. The behaviour 
of $l_{\mathrm{eff}}$ is simple. As long as the rupture grows as an 
expanding, roughly circular crack, $l_{\mathrm{eff}}$ scales with the 
rupture size itself. However, on shallow, elongated faults, the rupture 
cannot grow indefinitely in the vertical direction: it is confined by the 
seismogenic thickness. Once it spans the entire seismogenic layer, the 
crack ceases to expand and instead propagates along strike as a translating 
wave front - a self-healing slip pulse - whose width remains approximately 
constant, set by the seismogenic depth. At this point, $l_{\mathrm{eff}}$ 
saturates, regardless of how far the rupture travels along the fault.
	
	Continental crustal faults are typically locked below a certain depth 
	due to the brittle-ductile transition, where temperatures become 
	sufficiently high for rocks to deform plastically rather than 
	frictionally. The seismogenic depth, denoted $Z$, is the vertical extent of the locked, brittle portion of the fault. 
	For most continental faults, $Z \approx 10$-$20$ km, 
	varying with the geothermal gradient and rock composition.
	
	When an earthquake nucleates, the initial rupture expands as a crack. 
	However, once the crack's vertical extent reaches $Z$, 
	it can no longer grow in that direction. The rupture then becomes 
	channeled in the horizontal direction, propagating as a pulse 
	\cite{Heaton1990,Zheng1998,Ampuero2008}. The slip pulse is a traveling 
	wave of localized deformation: the fault slips only within a narrow 
	zone behind the rupture front, then heals behind the pulse, 
	so that the total slip at a given point is accumulated during a short 
	time window.
	
	The along-strike extent of the slipping zone in a pulse-like 
	rupture, denoted $L_{\mathrm{pulse}}$, is set by dynamic rupture 
	mechanics. Numerical simulations \cite{Ampuero2008,Gabriel2012} and 
	theoretical analyses \cite{Rice1983,Zheng1998} show that stable slip 
	pulses have a characteristic width that scales with the seismogenic 
	depth:
	\begin{equation}
		L_{\mathrm{pulse}} = \gamma \, Z,
		\label{eq:Lpulse}
	\end{equation}
	where $\gamma$ is a dimensionless pulse aspect ratio (typically $\gamma \approx 1$-$2$) depending on the rupture speed, the fault stress state and frictional properties \cite{Weng2017, Weng2019}. 
	
	Then, the effective structural scale that controls $D_c$ is $l_{\mathrm{eff}} = \min\!\big(L_{\mathrm{rup}}, \; \gamma \, Z\big)$. 
	For earthquakes with rupture lengths smaller than the pulse saturation 
	scale ($L_{\mathrm{rup}} < \gamma Z$), the event is 
	crack-like and $l_{\mathrm{eff}} \approx L_{\mathrm{rup}}$. For larger 
	events ($L_{\mathrm{rup}} \gg \gamma Z$), the rupture is 
	pulse-like and $l_{\mathrm{eff}}$ saturates at $\gamma Z$, 	which for typical crustal parameters is about $15$--$40$ km.
	
	Substituting $l_{\mathrm{eff}}$ into Eq.~(\ref{eq:DcGeom23}), we obtain 
	a more complete scale dependence of $D_c$:
	\begin{equation}
		D_c^{\mathrm{geom}} = \frac{C}{2\mu} \, l_{\mathrm{eff}}^{\zeta} = 
		\frac{C}{2\mu} \big[ \min(L_{\mathrm{rup}}, \gamma Z) 
		\big]^{\zeta}.
		\label{eq:DcComplete}
	\end{equation}
The theory thus predicts that $D_c$ grows sublinearly with rupture size 
while the rupture expands as a crack, and then saturates at a constant 
value of a few metres once the rupture spans the seismogenic layer and 
propagates as a steady pulse, irrespective of the total fault length.
	
    \section*{The role of cohesion, friction and geometric locking}
	\label{sec:cohesion}
	The discussion so far has treated fault as purely hierarchical interfaces: to unlock them, the system must slide up the bump, working 
	against the normal load. 
	
	However, real fault surfaces at depth are not 
	simply rough contacts in a frictional equilibrium. The contacts are 
	cemented by mineral precipitates, cold-welded by high normal stresses, 
	and possibly healed by chemical processes during the interseismic period. This introduces cohesion: a finite shear strength that must be 
	overcome even at zero normal stress \cite{Weiss2016}.
	Cohesion alters qualitatively the failure mode of small asperities \cite{Martel1997}. 
	Instead of sliding plastically over the geometric bump, a strongly bonded asperity may fail by brittle fracture: a crack propagates through the base of the bump, separating it from the substrate \cite{Hsu1969}. The slip required for fracture can be much smaller than the geometric unlocking distance under certain conditions, and it is controlled by the fracture process zone rather than by the bump geometry \cite{Tenthorey2006}.

\subsection*{Physical constraints on asperity failure}
An asperity of linear size $l$ bonded to the opposing fault face by 
cohesion $c$ can fail in different ways, depending on the dominant contribution to stability at different spatial scales. 

Indeed, three co-existing physical mechanisms contribute to the total resistance, namely cohesive strength, frictional resistance and geometric locking, each slightly prevailing over a different range of the spatial spectrum because they scale differently with $l$.

At the smallest scales, failure is governed by the fracture toughness 
$K_c$ of the bonded interface. For a crack whose length is comparable to the asperity size, the stress required for brittle failure is
\begin{equation}
	\tau_{\mathrm{frac}}(l) = \frac{\alpha K_c}{\sqrt{l}},
	\label{eq:tauFrac}
\end{equation}
where $\alpha \approx 1.12\sqrt{\pi/2}$ for a mode II edge crack 
\cite{Lawn1993}. Equation~(\ref{eq:tauFrac}) scales as $l^{-1/2}$: 
smaller bumps are harder to break. The slip distance associated with 
this process is the fracture process zone size,
\begin{equation}
	\delta_{\mathrm{frac}} \sim \frac{K_c^{2}}{\tau_p^{2}},
	\label{eq:deltaFrac}
\end{equation}
where $\tau_p$ is the peak strength of the bonded interface. For typical crustal rocks $\delta_{\mathrm{frac}} \sim 1$--$100~\mu$m, which coincides with the values of $D_c$ measured in laboratory friction experiments \cite{Dieterich1979,Marone1998}. At the bottom of this regime $D_c$ is roughly a material constant, independent of the size of the slipping patch.

At intermediate scales, all the three terms contribute, with friction becoming slightly dominant. The total sliding strength can be written as
\begin{equation}
	\tau_{\mathrm{slide}}(l) = c + \sigma_n\left( \mu(l) + C l^{\zeta-1} \right),
	\label{eq:tauSlideFull}
\end{equation}
where the three terms represent cohesive strength, frictional resistance, and the (dilatant) geometric interlocking, respectively. 

Although the friction coefficient $\mu$ is commonly treated as a constant, it is itself an emergent property of the fractal contact population \cite{Zaccagnino2025}. 
For a self-affine surface with fractal dimension $D$, the effective friction coefficient acquires the scale dependence 
\begin{equation}
	\mu(l) = \mu_0 \left( \frac{l}{l_0} \right)^{D-2},
	\label{eq:muScaling}
\end{equation}
where $D$ is the fractal dimension of the contact set, $l_0$ is a 
reference scale, and $\mu_0$ is the friction coefficient measured at 
that scale. In planar cross section, faulting usually shows $D \approx 1.5-1.7$ \cite{Hirata1989, Kagan1991, Cowie1995, BenZion2003}, giving $D-2 \approx -0.5 \div -0.3$. The frictional contribution to the sliding strength therefore scales as 
\begin{equation}
  \tau_{\mathrm{fric}}(l) = \sigma_n \mu(l) \propto l^{-2+D}, 
\end{equation}
which is close to, but usually slightly shallower than, the $l^{-1/2}$ scaling of cohesive fracture. The critical slip distance in this transitional regime is no longer constant but begins to increase, reflecting the growing real contact area.

At the largest scales, the controlling process is dominated by geometric unlocking of asperities. 
The dilatant term in Eq.~(\ref{eq:tauSlideFull}) is
\begin{equation}
	\tau_{\mathrm{geom}}(l) = \sigma_n C l^{\zeta-1},
	\label{eq:tauGeom}
\end{equation}
which, for $\zeta = 2/3$, scales as $l^{-1/3}$, the shallowest of the 
three exponents. Therefore, geometric unlocking of asperities controls the stress level - slightly overcoming fracture and friction terms - governing the failure of the largest asperities, hence earthquake dynamics at the largest scales. 

The associated critical slip distance is
\begin{equation}
	D_c^{\mathrm{geom}}(l) = \frac{C}{2\mu} \, l^{\zeta},
	\label{eq:DcGeomFinal}
\end{equation}
which increases with the size of the bump and becomes a structural, rather than a material, property of the fault.

The three contributions -- cohesive fracture ($\sim l^{-1/2}$), frictional resistance ($\sim l^{-0.4}$), and geometric unlocking ($\sim l^{-1/3}$) -- are simultaneously present over the entire spectrum of scales, but their relative importance shifts continuously with $l$ because of their different scaling exponents. 
Since these exponents are close to one another, the crossovers are not sharp: fracture and friction overlap over a broad interval, as do 
friction and geometric unlocking. This continuous transition is a direct consequence of the fault self-affine architecture.

This perspective is in agreement with the evidence that big events involve major existing faults producing frictional slip along them and with widespread ruptures at local scale.

Notice that the mesoscale frictional regime and the macroscale geometric locking phase can be interpreted as emergent properties ruled by the same physics, where friction corresponds to structural pinning in the limit of several small asperities. 

The effective critical slip distance is the largest among the competing length scales. Introducing, as in the previous section, the effective structural length $l_{\mathrm{eff}} = \min(L_{\mathrm{rup}}, \gamma Z)$, which accounts for both the crack-like expansion of small ruptures and the pulse saturation of large ones, we obtain the unified expression
\begin{equation}
		D_c(l_{\mathrm{eff}}) = \max\!\left( \delta_{\mathrm{frac}}, \; 
		\frac{C}{2\mu} \, l_{\mathrm{eff}}^{\zeta} \right),\label{eq:DcMaxUnified}
\end{equation}
where $\zeta \simeq \zeta_{\text{KPZ}} = 2/3$ is the roughness exponent, $C$ and $\mu$ are measurable fault properties, $\delta_{\mathrm{frac}}$ is constrained by rock mechanics, and $l_{\mathrm{eff}}$ is set by the rupture length and the seismogenic depth $Z$. 

The frictional contribution, which governs the transition between the cohesive floor and the geometric branch, is absorbed into the crossover behaviour. Below the crossover scale, cohesive failure dominates and $D_c$ is the material constant $\delta_{\mathrm{frac}}$, as observed in the laboratory. Above the 
crossover, geometric unlocking dominates and $D_c$ grows as 
$l_{\mathrm{eff}}^{2/3}$, becoming a structural property that saturates at a few metres for the largest earthquakes, when $l_{\mathrm{eff}}$ reaches the pulse width $\gamma Z$.

\begin{figure*}
	\includegraphics[width=2\columnwidth]{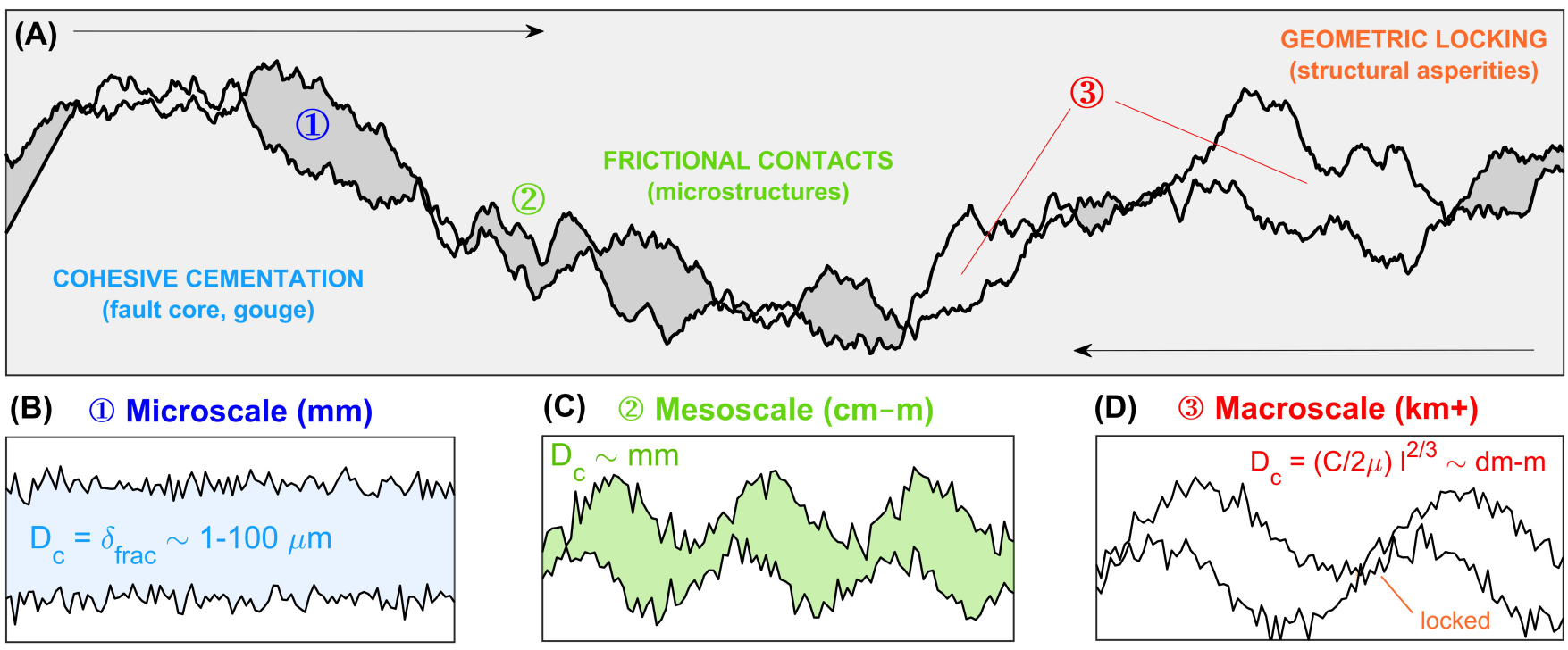}
	\caption{Two interlocking rough fault surfaces (A) with locked zones (contacts) alternated with unlocked segments shaded in gray - in nature and in laboratory observations they are filled with unconsolidated gouge, finely damaged-pulverized rocks or cemented materials. Three magnification windows illustrate the dominant failure mechanism at each scale. B) at the microscale, cohesive mineral cement bridges the gap and the critical slip distance is the fracture process zone, at this scale $D_c = \delta_{\mathrm{frac}} \sim \mu$m. C) at the mesoscale, frictional point contacts appear while some cohesive bonds still contribute to fault stability. D) At the macroscale (km), geometric interlocking dominates with multiple overlaps; the unlocking slip grows with bump size as $D_c = (C/2\mu)\,l^{2/3} \sim$ m.}
	\label{fig:fricfrac}
\end{figure*}

\begin{figure*}
	\includegraphics[width=2\columnwidth]{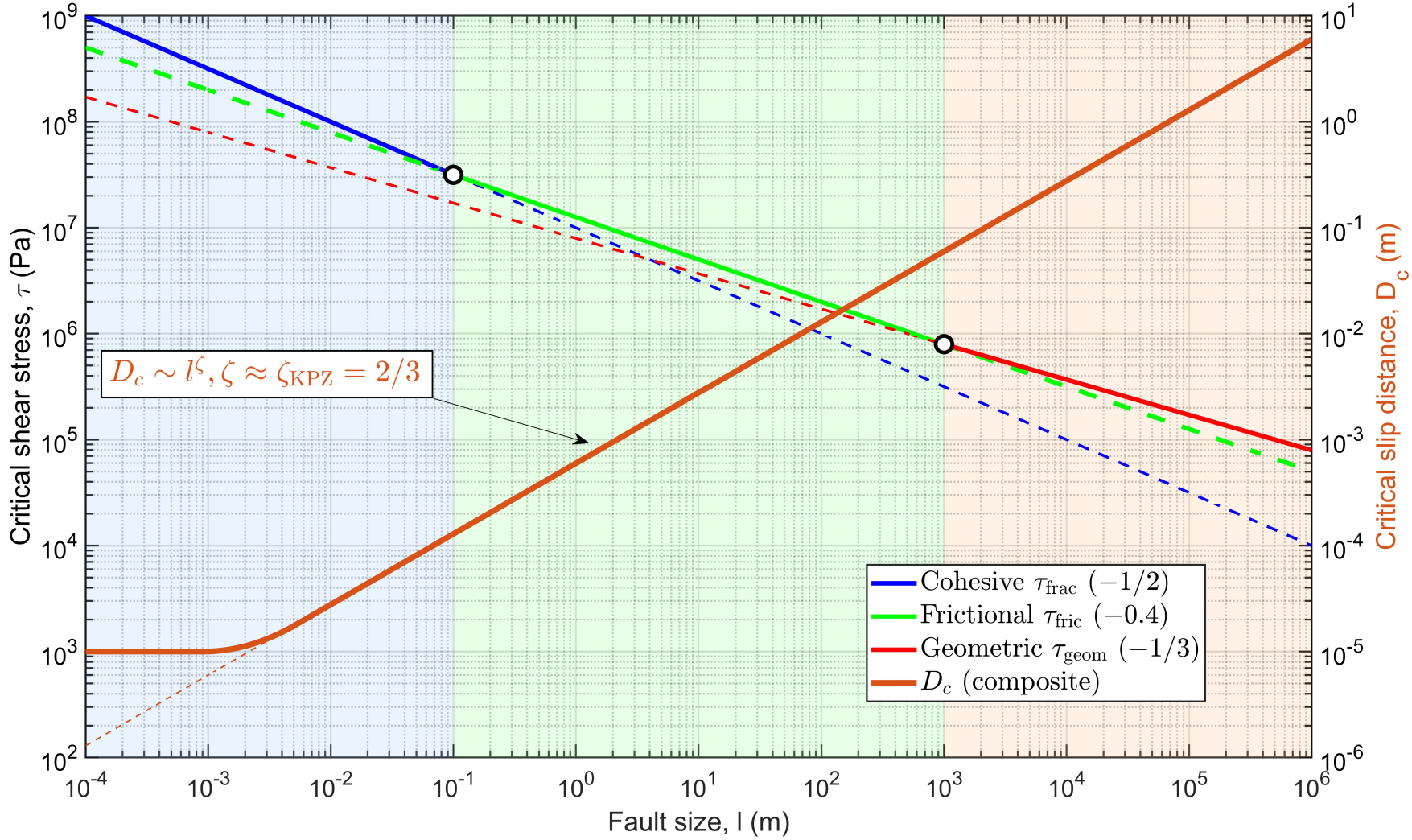}
	\caption{
			Failure mechanism map and critical slip distance.
			Three physical contributions to the shear strength of a rough fault are shown as functions of asperity size $l$: cohesive fracture ($\sim l^{-1/2}$, blue), frictional resistance ($\sim l^{-0.4}$, green), and geometric unlocking ($\sim l^{-1/3}$, red).
			Each mechanism is drawn solid where it dominates and dashed elsewhere, with indicative crossovers (white circles).
			The background shading marks the three regimes: cohesive (blue), frictional (green), and geometric (orange).
			The composite critical slip distance $D_c$ (thick black curve, right axis) transitions smoothly from the constant cohesive floor $\delta_{\mathrm{frac}} \sim 5~\mu\mathrm{m}$ to the scale-dependent geometric trend $D_c \propto l^{2/3}$, with the departure occurring at or below the laboratory scales.
			The proximity of the three scaling exponents ($-0.5$, $-0.4$, $-0.3$) ensures that the crossovers are broad, so that the three mechanisms overlap over a substantial range of scales.}
	\label{fig:scheme}
\end{figure*}

See Figure~\ref{fig:fricfrac} for a visual representation and Figure~\ref{fig:scheme}. 

\section*{An application: the case of the 2025 Myanmar Earthquake}
\label{sec:Myanmar}
On 28 March 2025, a $M_w$ 7.7-7.8 earthquake struck central Myanmar, 
rupturing a $\sim 450$~km segment of the Sagaing Fault, a major 
right-lateral strike-slip plate boundary between the Indian and Sunda 
plates \cite{Latour2025}. 

This event is known for providing the first direct visual inversion of the critical slip distance at surface, yielding a value of $D_c \approx 3$~m. 

In light of this special, unprecedented observational novelty, this event provides an ideal test of our theory: the fault is a long, mature 
strike-slip system with well-developed self-affine roughness; the rupture 
length far exceeds the seismogenic depth, placing it firmly in the 
pulse-saturated regime; and the inferred $D_c$ of $\sim 3$~m is typical 
of large crustal earthquakes, neither anomalously small nor large.

We can apply the unified expression, Eq.~(\ref{eq:DcMaxUnified}), using 
reasonably constrained parameters: for the roughness amplitude, we 
adopt $C \approx 3 \times 10^{-3}~\mathrm{m}^{1/3}$, a value within the 
typical range for natural faults and consistent with observations \cite{Candela2009,Brodsky2011}. 
The friction coefficient is taken as $\mu = 0.2$, a value compatible with the resistance to slip of intraplate continental mature faults. $Z \approx 10$~km, based on the depth extent provided by most slip inversions. The pulse aspect ratio is set to $\gamma \approx 1$ a value compatible with dynamic rupture simulations \cite{Ampuero2008,Gabriel2012}. The rupture length is $L_{\mathrm{rup}} \approx 450$~km, as estimated from aftershock 
distributions and finite-fault inversions. The effective structural 
length is therefore $l_{\mathrm{eff}} = \min(450~\text{km}, 1.0 \times 
10~\text{km}) = 10$~km. Since $l_{\mathrm{eff}} \gg l^*$, the event 
lies in the geometric unlocking regime, and the cohesive floor 
$\delta_{\mathrm{frac}}$ is completely negligible.

With these values, the predicted critical slip distance is
\begin{equation}
	D_c^{\mathrm{pred}} = \frac{C}{2\mu} \, l_{\mathrm{eff}}^{2/3} 
	= \frac{3 \times 10^{-3}}{2 \times 0.2} \times 10^{8/3} \approx 3.5~\text{m},
\end{equation}
in good agreement with the observed value of $\sim 3$~m.

This prediction uses only (potentially) independently measurable fault properties--roughness amplitude, friction coefficient, and seismogenic depth--without any tuning or hidden parameters. 

The laboratory-measured cohesive floor $\delta_{\mathrm{frac}} \sim 10$-$100~\mu\mathrm{m}$ is entirely negligible at this scale, contributing less than $0.01\%$ to the total $D_c$ and well above the prediction of few centimeters for the grain sizes at seismogenetic depth by classical models (e.g., \cite{Scholz1988}). 

The 2025 Myanmar earthquake thus provides strong quantitative support for 
our geometric theory. The observed $D_c$ of $\sim 3$~m is precisely what 
one expects from the roughness amplitude of a mature fault and the pulse 
saturation width of a dozen kilometers. This reinforces the central 
message of our framework: $D_c$ for large earthquakes is a structural 
property of the fault, governed by its multi-scale geometry, not an intrinsic material friction parameter.
	
\section*{Earthquakes as hierarchical asperity cascades}
\label{sec:cascade}
The unified expression for $D_c$, Eq.~(\ref{eq:DcMaxUnified}), identifies the critical slip distance as a local property of the largest bump that must be unlocked within the slipping patch. However, it does not tell us what determines the size of that segment. To answer this question we must consider the fault not as a single asperity but as a statistical population of barriers distributed over a wise range of scales according to the self-affine roughness described in the previous sections. 

A growing rupture interacts with this population sequentially (at least before it enters the pulse-like saturating regime, that here we neglect): it may unlock a bump of size $l$, only to be blocked by a steeper asperity at a slightly larger scale. Whether the cascade continues or arrests is determined by the extreme value statistics of the roughness slope distribution. 

This section develops the probabilistic description of that cascade and derives the resulting earthquake size distribution with implications for the relationship between the standard rate and state framework and our new geometric structural formalism. 

To describe the asperity population statistics, we discretize the 
fault into a hierarchy of patches. A patch of linear dimension $l$ 
contains a characteristic bump of height $w(l) = C l^\zeta$ and has 
a shear strength $\Delta\tau(l)$ which is a decreasing function of $l$: large bumps are less steep and therefore contribute proportionally less to the total strength.

The local slope of the interface at scale $l$ is $S_l(x) = [h(x+l)-h(x)]/l$, a Gaussian random variable with zero mean and variance	$\sigma_S^2(l) = C^2 l^{2\zeta-2} = C^2 l^{-2/3}$, inheriting its statistics from the KPZ equation. A bump arrests the cascade if its local maximum slope $S_{\max}$ exceeds the critical value
\begin{equation}
	s_c = \frac{\Delta\tau}{\sigma_n}.
	\label{eq:criticalSlope}
\end{equation}
The probability that a window of length $l$ contains such an arresting bump is an extreme value problem that can be treated analytically under suitable hypotheses (see the supplementary materials for details).

The expected number of arresting bumps $\Lambda$ at scale $l$ can be written as 
\begin{equation}
	\Lambda(l) = \frac{L_{\max}}{l} \, \mathbb{P}\!\left( S_{\max}(l) > s_c \right); 
	\label{eq:hazardRate}
\end{equation}
then, treating the cascade as a Poisson process in scale space, the probability that the rupture reaches size $L$ without encountering an arresting bump at any smaller scale reads
\begin{equation}
	P_{\mathrm{surv}}(L) = \exp\!\left[ - \int_{l_{\min}}^{L} \Lambda(l) \, \frac{dl}{l} \right].
	\label{eq:survivalIntegral}
\end{equation}
The integral is dominated by the upper limit because the integrand decays exponentially (see Supplementary Material for the full derivation). The leading-order result is a Weibull survival function
\begin{equation}
	P_{\mathrm{surv}}(L) \approx \exp\!\left[ - k \left( \frac{L}{L_{\mathrm{crit}}} \right)^{\beta} \right],
	\label{eq:WeibullForm1}
\end{equation}
with $L_{\mathrm{crit}} = \left( \frac{\sigma_n C}{\Delta\tau} \right)^3$, $\beta = 1 - \zeta = 1/3$ and $k$ is a dimensionless constant.  

Equation~(\ref{eq:WeibullForm1}) is a Weibull survival function: for $L \ll L_{\mathrm{crit}}$ it is well approximated by a power-law, while for
$L \gg L_{\mathrm{crit}}$ the exponential cutoff suppresses the
probability of ruptures that exceed the critical scale.

The probability density of the final rupture length - its negative derivative with respect to the length itself - for $\beta = 1/3$ gives $p(L) \propto L^{-2/3}$ for $L \ll L_{\mathrm{crit}}$.  

Converting this to a magnitude-frequency distribution requires assumptions about the moment-area scaling and the distribution of $L_{\mathrm{crit}}$, which prevent us to provide a straightforward derivation of the Gutenberg-Richter law with a $b$-value $= 1$ as observed. 

The connection between the two remains an open question that likely involves the statistics of $L_{\mathrm{crit}}$ over a heterogeneous fault population.
Indeed, the critical scale $L_{\mathrm{crit}} = (\sigma_n C / \Delta\tau)^3$ depends on the ratio of roughness amplitude to stress drop: rougher faults or those with lower stress drop have larger $L_{\mathrm{crit}}$, implying that larger events are more probable in this configuration because stronger barriers can only be overcome by more energetic ruptures. This may provide a natural upper bound on earthquake size that emerges from the roughness statistics rather than being imposed empirically. However, this research is beyond the scope of this article. 

Conversely, an important conceptual point to discuss here is the relationship between the nucleation scale and the final earthquake size.  In the cascade model, an earthquake can nucleate on any asperity, but the final size is determined by the largest bump that the cascade can unlock before arresting or eventually evolving into a traveling pulse.

The critical slip distance measured for the event, $D_c^{\text{(obs)}}$, is therefore the
geometric unlocking distance of that dominant bump,
\begin{equation}
	D_c^{\text{(obs)}} = \frac{C}{2\mu} \, L_{\text{dom}}^{\zeta}.
	\label{eq:eventDc}
\end{equation}  
As a consequence, $D_c$ is not a property of the nucleation zone but of the largest asperity the earthquake breaks.  
Small and moderate earthquakes, for which $L_{\text{dom}} \lesssim \gamma Z$, have $D_c$ that scales with their size; large
earthquakes have $D_c$ saturated at the pulse width, explaining why
$D_c$ inferred from seismology correlates with magnitude for moderate
events but it saturates at larger sizes.  
On the other side of the spatial spectrum, for tiny events below the crossover scale $l^*$, the barrier strength is dominated by cohesion and is roughly independent of scale, leading to a more characteristic size distribution near the grain-size cutoff.

In the next section we show how the same fractal contact population that governs the cascade can be used to derive standard rate-and-state
friction laws as its mean-field limit, allowing us to reconcile our formalism with the standard scale-free ones. 
	
\section*{Emergence of rate-and-state friction from fractal contact dynamics}
\label{sec:RSemergence}
We now show that the standard rate-and-state friction equations, long
regarded as phenomenological constitutive laws describing the macroscopic sliding behaviour of faults, are the mean-field signature
of a population of discrete, fractal-distributed asperity contacts. Specifically, their mathematical form does not encode any intrinsic material property of the interface; rather, it emerges from the statistics of the underlying contact population. The parameters $a$, $b$, and $D_c$ of the rate-and-state friction laws (Equation~\ref{eq:RSintro}) are therefore not constants depending on the material, but collective properties that depend on the fault's geometric structure and on the scale of the slipping patch. This distinction is essential: a constitutive law prescribes how a material element responds to imposed conditions independently of the system size, whereas the equations we
derive here acquire their form and their parameter values from the
multiscale roughness of the specific fault under consideration.

The derivation provided in details in the supplementary materials gives a mechanistic description for the empirical framework discussed so far and reveals that its parameters are not independent material constants but, once again, collective properties of fault's geometric structure.

Consider a fault as a collection of discrete asperity contacts whose sizes follow the power-law distribution $n(l) \propto l^{-4/3}$ inherited from the self-affine roughness. Each contact has a finite lifetime: it breaks when the accumulated slip reaches the critical distance $d_c(l) = \max(\delta_{\mathrm{frac}}, \frac{C}{2\mu} l^{\zeta})$, after
which it is replaced by a new contact of the same size. The steady-state age distribution of the contact population follows from the McKendrick-von Foerster equation \cite{Mckendrick1925, VonFoerster1959} and takes the exponential form $N_0(t_c) \propto e^{-t_c/t_{\rm life}}$, where $t_{\rm life} = d_c/V$ is the average contact lifetime at slip velocity $V$ and $t_c$ is the time elapsed since contact formation.

Assuming the shear strength of a contact grows logarithmically with its age, as widely observed in laboratory experiments \cite{Marone1998}, the macroscopic shear stress is obtained by averaging the forces sustained by all contacts over the fault area. Each contact of size $l$ has an area $A_c(l) \propto l^2$, so that larger contacts contribute disproportionately to the total frictional resistance. This ensemble average naturally separates into three contributions (see Supplementary Material for the full derivation): a size-averaged instantaneous strength, a term proportional to $-\ln V$ that represents the direct velocity effect, and a term involving the contact-area-weighted average of $\ln d_c(l)$ that represents the state contribution.

Comparing these terms with the standard rate-and-state formulation $\tau = \sigma_n [\mu_0 + a \ln(V/V_0) + b \ln(V_0 \theta / D_c)]$, we identify the dimensionless rate-and-state parameter $a$ as
\begin{equation}
	a = \frac{\bar{a} \rho_A}{\sigma_n},
	\label{eq:abIdentification}
\end{equation}
where $\bar{a}$ (units of stress) is the logarithmic strengthening rate of a single contact defined by the age-dependent contact strength (see the supplementary materials). $\rho_A = \int A_c(l) N_{\mathrm{tot}}(l) dl$ is the total contact area per unit fault area, a dimensionless measure of the real area of contact.

Equation~(\ref{eq:abIdentification}) shows that the direct effect coefficient $a$ is not a material property, but reflects the microscopic contact-scale strengthening $\bar{a}$, the real contact area $\rho_A$, and the ambient normal stress. 

Empirically, most experiments find $b \approx a$, giving nearly velocity-neutral friction at steady state once all the uncertainties are considered, also consistent with observations on mature faults.
Hence, at least for theoretical purposes, we do not go into details for the derivation of the parameter $b$ of rate-and-state friction laws assuming, as a first-order approximation, that it follows the same scaling as $a$.

Regarding $D_c$, the critical slip distance that appears in the macroscopic law is obtained by equating the state contributions from our microscopic derivation with those of the standard rate-and-state formulation. The result is a weighted geometric mean of the critical distances $d_c(l)$ across the whole spectrum of scales involved in the contact population:
\begin{equation}
	D_c \approx d_0 \exp\!\left[\frac{\int_{l_{\min}}^{L_{\max}} A_c(l) N_{\mathrm{tot}}(l) \ln(d_c(l)/d_0) dl}
		{\int_{l_{\min}}^{L_{\max}} A_c(l) N_{\mathrm{tot}}(l) dl} \right],
	\label{eq:DcGeneralt}
\end{equation}
where we approximated a factor $\frac{a}{b} \approx 1$, while $d_c(l) = \max(\delta_{\mathrm{frac}}, \frac{C}{2\mu} l^{\zeta})$ is the critical slip distance of a single contact of size $l$, $A_c(l) \propto l^2$ is the area of that contact,
$N_{\mathrm{tot}}(l) \propto l^{-4/3}$ is the number of contacts of size $l$ per unit fault area per unit size, and $a$ and $b$ are the dimensionless rate-and-state parameters identified in Eq.~(\ref{eq:abIdentification}). $d_0$ represents our arbitrary choice of reference length. Equation~(\ref{eq:DcGeneralt}) is the
central result of this section: it expresses the macroscopic critical slip distance as a structural average over the fault's multiscale roughness.

Mathematically, $D_c$ is the geometric mean of the size-dependent critical slip distances
$d_c(l)$, weighted by the contact area distribution
$A_c(l) N_{\mathrm{tot}}(l)$.

The weighting by contact area $A_c(l)$ is essential: larger contacts, which control the geometric locking, dominate the average despite being far less numerous than the small, stronger, cohesion-dominated contacts.

The physical content of Eq.~(\ref{eq:DcGeneralt}) depends on the domain of integration. 

If the average is taken over all contacts present on the fault, the strong divergence of the integrand $A_c N_{\mathrm{tot}} \propto l^{2/3}$ at large $l$ weights the result toward the largest bumps in the slipping patch. Restricting the integration to scales up to
$l_{\mathrm{eff}} = \min(L_{\mathrm{rup}}, \gamma Z)$ -- the size of the largest bump that can be unlocked by a given earthquake -- gives
\begin{equation}
	D_c \approx \frac{C}{2\mu} \, l_{\mathrm{eff}}^{2/3},
	\label{eq:DcEmergent}
\end{equation}
recovering the geometric scaling derived from the single-asperity unlocking model. 

Conversely, the floor $\delta_{\mathrm{frac}}$ is recovered when $l_{\mathrm{eff}} \lesssim l^*$, i.e., when the entire slipping patch is smaller than the fracture-sliding crossover scale.

This result has three important implications. 

First, rate-and-state friction is an emergent phenomenology, not a fundamental constitutive law: its mathematical form arises from statistical averaging over a fractal contact population, and its parameters -- including $a$, $b$, and $D_c$ -- are structural properties of the fault, not intrinsic material constants. At the same time, this emergent origin explains the remarkable success of the rate-and-state framework in earthquake science: because the underlying contact dynamics are largely universal,
the same mathematical structure appears to correctly reproduce the behavior of a wide range of materials and conditions, giving the framework a robustness that has made it the standard tool of earthquake mechanics for decades. Our derivation thus does not invalidate rate-and-state friction but rather provides it with a deeper physical foundation, anchoring its parameters in the measurable geometry of the fault surface. However, it also makes clear why laboratory values of $D_c$ (as well as $a$ and $b$) cannot be extrapolated directly to natural faults: the contacts that dominate the average in a centimetre-scale sample are not the same as those that control weakening in a kilometre-scale rupture, so the effective $D_c$ is inherently scale-dependent.

Second, the same theory predicts $D_c \sim \mu\mathrm{m}$ for the smallest
subcentimetre-scale laboratory samples (where the largest contacts are
sub-$l^*$) and $D_c \sim \mathrm{m}$ for kilometre-scale ruptures (where
geometric unlocking dominates), explaining the observed upscaling by several
orders of magnitude without invoking any scale-dependent material
properties. 

Third, the apparent constancy of $D_c$ in some laboratory settings is a
consequence of the fact that those laboratory samples probe only the small-scale, cohesion-dominated end of the contact spectrum; the true scale dependence may become visible only when larger fault patches are activated.

When the slip velocity changes, the contact population relaxes toward a
new steady state on a time scale $D_c/V$, from which the Dieterich aging
law $\dot{\theta} = 1 - V\theta/D_c$ follows directly (see Supplementary
Material). 

The rate-and-state framework, which has been regarded as a phenomenological description of rock friction for a long time, can then find its physical justification in the hierarchical geometry of the fault surface.

\section*{Discussions}
\label{sec:discussion}
\subsection*{Reinterpretation of rate-and-state friction}
Our results demand a fundamental reinterpretation of the rate-and-state 
friction framework. 

The parameters $a$, $b$, and $D_c$ are not independent material constants to be fitted from each experiment, but are linked through the underlying fault roughness. 

The empirical success of rate-and-state friction in modelling laboratory experiments is not evidence for the intrinsic nature of $d_0$ (and so $D_c$), but rather a consequence of the fact that laboratory samples are too small to enter the geometric scaling regime given their uncertainties. The laboratory $d_0$ is both a proxy of the fracture process zone size, a material property that controls cohesive failure at the smallest scales, and experimental constrains and assumptions. 
Such quantities are replaced by geometric locking in earthquakes; therefore, the critical slip distance in earthquakes nothing has to share with the deemed small-scale equivalent measurements in the laboratory.  
Rate-and-state friction, rather than being a fundamental 
constitutive description, emerges as the mean-field limit of the 
underlying fractal contact dynamics: its parameters are collective 
properties of the fault's geometric structure, not independent scale-invariant inputs. Their values cannot be directly extracted from the lab to natural fault systems, but require appropriate upscaling. 

\subsection*{Towards a multi-scale characterization of fault stability}
We have available precise terms for the conditions under which a rupture begins - strength, friction, toughness - nevertheless, there exists no term for the property that determines whether a rupture, once initiated, will stop or 
grow from small to large scales. 
This distinction is actually fundamental: a fault may be weak and easy to start ruptures; nevertheless, it can be highly resistant to propagation; or conversely, it may  be strong and hard to start, but, once overcome, offer little resistance to unbounded growth.

Here, we introduce the concept of \textquotedblleft fault retentivity'', denoted $\mathcal{R}(l)$ to express this concept.  

We define it as the multiscale structural capacity of a fault to arrest a propagating rupture through the hierarchical population of geometric, cohesive, and frictional barriers embedded in its self-affine topography.

Fault retentivity can be mathematically expressed as 
\begin{align}
	\mathcal{R}(L) &= 1 - P_{\mathrm{surv}}(L) \nonumber \\
	&= 1 - \exp\!\left[ - \int_{l_{\min}}^{L} \frac{L_{\max}}{l} \; 
	\mathbb{P}\!\left( S_{\max}(l) > \frac{\Delta\tau}{\sigma_n} \right) 
	\frac{dl}{l} \right], 
	\label{eq:retentivity}
\end{align}
which gives the probability that a nucleated rupture is stopped before it reaches size $L$.  

A fault with high retentivity arrests most ruptures at small scales: $\mathcal{R}(L)$ rises rapidly toward unity at small $L$, suppressing large events.  

A fault with low retentivity allows cascades: $\mathcal{R}(L)$ remains close to zero until large $L$, so occasional great earthquakes can occur.  

Conceptually, this term unifies the observable features of fault behaviour -- $b$-value, maximum magnitude, and effective $D_c$ -- as different aspects of the same underlying arrest statistics.

It is also different from existing concepts.
Indeed, while strength governs nucleation; retentivity governs arrest.
Toughness is a local material property; retentivity is an emergent, scale-dependent statistical property of the entire fault structure. 
Seismic coupling describes the partitioning of slip between seismic and aseismic fault slip modes; retentivity describes the size distribution of the seismic events that do occur. 
The $b$-value is an empirical descriptor of that distribution; retentivity 
is its physical cause. 

Together with strength, friction, toughness, and seismic coupling, retentivity completes the vocabulary needed to describe the mechanical state of a fault and provides a theoretical physical background justification to the $b$-value spatial variations.

\subsection*{Implications for rupture arrest and seismic hazard}
The hierarchical cascade model provides a natural explanation for 
earthquake propagation and arrest. 
The initial slip instability may begin on a very small asperity, where $D_c$ is tiny. As the rupture grows, it must unlock progressively larger bumps, each requiring larger slip. The effective $D_c$ for the event is set by the largest dominant bump that must be overcome, implying that the nucleation zone size grows with the eventual earthquake magnitude. 

Potentially, this may open the possibility of predicting $D_c$, and hence 
fracture energy, before an earthquake occurs, by measuring fault 
roughness, with implications for seismic potential assessment \cite{VenegasAravena2024, VenegasAravena2025}. Combined with estimates of stress drop and seismogenic depth, this would allow physics-based hazard estimates that do not rely on empirical scaling relations; moreover, it may help explain the occurrence of highly variable fault response to stress and strain during the seismic cycle \cite{Caniven2017, Dolan2017, Zaccagnino2021, Barbot2023, Brodsky2026, Zaccagnino2026b}.

\subsection*{Limitations and future work}
The present theory rests on several simplifying assumptions that deserve further investigation: we have treated the fault as a one-dimensional profile with Gaussian height statistics; however, real faults are roughly two-dimensional surfaces with anisotropic, possibly non-Gaussian roughness. 

Moreover, our cascade model assumes quasi-static failure of each scale before loading the next, while real ruptures are dynamic and may even jump over barriers. 

We have also assumed that large earthquakes consist of a single traveling pulse (an hypothesis with strong theoretical and observational background \cite{Heaton1990, Melgar2017, Lambert2021}), and that fault roughness is static over the seismic cycle. 

Extending the theory to two dimensions, incorporating dynamic effects lubricating faults \cite{Brodsky2001, DePaola2011, Pozzi2021}, and coupling roughness evolution with the seismic cycle represent natural directions for our future theoretical work.

\section*{Conclusions}
We have presented a first-principles theory for the critical slip 
distance $D_c$ on rough faults, from laboratory to tectonic scales. 

We reached some important results. 

First, $D_c$ is not a material constant but a structural property of the fault surface. 
It is governed by the self-affine roughness amplitude $C$, the 
roughening (Hurst) exponent $\zeta \approx 2/3$, and the friction coefficient $\mu$, through the unified expression 
$D_c = \max(\delta_{\mathrm{frac}}, \frac{C}{2\mu} l_{\mathrm{eff}}^{2/3})$, where $\delta_{\mathrm{frac}} \sim 1$--$100~\mu\mathrm{m}$ is the fracture process zone that sets the laboratory values, and $l_{\mathrm{eff}} = \min(L_{\mathrm{rup}}, \gamma Z)$ is the effective structural scale that saturates at the seismogenic depth for large earthquakes, limiting $D_c$ at a few metres. The several-order-of-magnitude upscaling of $D_c$ from the laboratory to the field is thus explained by a single, parameter-free expression whose ingredients are potentially measurable.

Second, rate-and-state friction emerges from our theory as the mean-field limit of the underlying fractal contact dynamics. Its parameters are not intrinsic material constants but collective properties of the fault's geometric structure, and its mathematical form arises from the same self-affine roughness that governs $D_c$. This provides rate-and-state friction with a first-principles physical foundation, while simultaneously explaining why its parameters cannot be extrapolated from the laboratory to the field.

Third, the hierarchical cascade model naturally leads us to introduce the concept of \textquotedblleft fault retentivity'': the probability that a nucleated rupture is arrested before it reaches a given size, determined entirely by the multiscale population of barriers encoded in the fault's roughness. Retentivity is a key property that unifies the $b$-value, the maximum magnitude, and the effective $D_c$ as emergent consequences of the same underlying structure. It provides a single mathematical object -- the arrest probability as a function of scale -- that characterises the seismic 
potential of a fault from its geometric structure alone.

Fourth, The theory successfully predicts the observed $D_c \approx 3$~m for the 2025 $M_w 7.8$ Myanmar earthquake. This validates the claim that $D_c$ for large earthquakes can be potentially determined before an event occurs, by measuring fault's roughness.

Fifth, the broader and final implication is that assessing fault stability -- the propensity of a fault to host large or small earthquakes -- requires moving beyond the standard frictional framework. 
The quantity $a-b$ as a measure of fault stability is a theoretical oversimplification (although very simple and powerful first-order frictional estimator). It may be acceptable for laboratory studies, but it should not be applied to predict the slip behavior on natural faults. Indeed, the characterization of fault stability demands full consideration of the hierarchical architecture of the fault surface, which governs the entire spectrum of seismic behaviour through the retentivity function. 

Our framework eliminates hidden parameters from earthquake constitutive laws and replaces them with quantities that are directly measurable from fault structure, rock mechanics, and seismological observations, opening the possibility of future physics-based seismic hazard assessment grounded in fault structural properties.
	
\begin{acknowledgments}
   The authors thank Xiaofei Chen, Carlo Doglioni, Chris Marone, Haoran Meng, Paul A. Selvadurai, Didier Sornette, Filippos Vallianatos and J\'{e}r\^{o}me Weiss for fruitful discussions. 
   This research is supported by the Outstanding Oversea Postdoctoral Fund awarded to Davide Zaccagnino (Risks-X, SUSTech) within the Guangdong province Oversea Young Talents Program.  
\end{acknowledgments}

\newpage 
\appendix 
\onecolumngrid
\section*{SUPPLEMENTARY MATERIALS}
\section{Derivation of the hierarchical evolution of asperity cascades}
\label{sec:cascade2}

The number density of bumps of size $l$ along a fault profile 
follows from its fractal geometry. 

Representing the fault as a 1D self-affine curve with roughness exponent $\zeta$, the number of asperities of size $l$ covering the total fault length $L_{\max}$ scales as 
$(L_{\max}/l)^{D}$, where $D = 2 - \zeta$ is the fractal dimension of the profile. The number of bumps per unit fault length per unit size is therefore given by 
\begin{equation}
	n(l) = \frac{dN}{dl \, dL_{\mathrm{fault}}} = \frac{A}{l^{2-\zeta}} 
	= \frac{A}{l^{4/3}},
	\label{eq:sizeDist}
\end{equation}
for $l_{\min} \le l \le L_{\max}$, where $l_{\min}$ is the 
grain-size cutoff and $A$ is a normalization constant.

An earthquake begins when a small patch of size $l_0$ reaches its peak strength and fails dynamically. The stress drop $\Delta\tau$ of this initial failure loads the surrounding, larger asperities. If the stress increase is sufficient to bring a neighboring larger bump to its failure threshold, that bump also breaks, adding its slip to the rupture and further loading even larger bumps.                                    

A cascade emerges when a rupture that has grown to size $l$ 
unlocks the next, larger bump only if the stress drop it releases 
exceeds the excess strength of that bump. For bumps small enough to be 
in the fracture regime, this excess includes the cohesion that must be 
shattered; for larger bumps in the sliding regime, only the geometric 
interlocking remains. The rupture continues to grow as long as its stress drop outweighs the barrier it faces, and it stops when it encounters a bump too strong to break.

The barrier strength at a given scale is not deterministic, because 
the local slope fluctuates from one realization to another. The fault roughness is a random field, and the maximum slope within a window of size $l$ is a random variable drawn from the extreme value distribution of the underlying slope field.
The local slope of the interface at scale $l$ in the position $x$ is defined as
\begin{equation}
	S_l(x) = \frac{h(x+l) - h(x)}{l},
	\label{eq:localSlope}
\end{equation}
which is a Gaussian random variable with zero mean and variance
\begin{equation}
	\sigma_S^2(l) = \frac{\langle [h(x+l)-h(x)]^2 \rangle}{l^2} 
	= C^2 l^{2\zeta-2} = C^2 l^{-2/3}.
	\label{eq:slopeVariance}
\end{equation}
Indeed, the Gaussianity follows from the KPZ dynamics introduced in the main text: the fault surface evolves under stochastic 
growth driven by Gaussian white noise by hypothesis, and the resulting height field $h(x)$ is a Gaussian random field. Since the slope $S_l(x)$ is a linear combination of Gaussian variables, it inherits the same statistics.

The condition for a bump to arrest the cascade is that its local maximum slope exceeds the critical slope $s_c = \frac{\Delta\tau}{\sigma_n}$.

The probability that a single window of length $l$ contains a slope 
exceeding $s_c$ is an extreme value problem (i.e., peak over threshold).

For a stationary Gaussian process, the maximum over a segment of length $l$ which contains $N_{\mathrm{ind}}$ independent samples has a cumulative distribution that tends to a Gumbel 
distribution for large $l$ as stated by the Fisher--Tippett--Gnedenko Theorem \cite{Gnedenko1968}. 

Specifically, $\lambda_{\min}$, the correlation length, controls the number of independent samples as $N_{\mathrm{ind}} \approx l/\lambda_{\min}$. 

The exact probability that at least one sample exceeds $s_c$ is
\begin{equation}
	\mathbb{P}(S_{\max}(l) > s_c) = 1 - \left[ 1 - \bar{\Phi}\!\left( 
	\frac{s_c}{\sigma_S(l)} \right) \right]^{N_{\mathrm{ind}}},
\end{equation}
where $\bar{\Phi}(x) = \frac{1}{\sqrt{2\pi}} \int_x^\infty e^{-u^2/2} du$ is the complementary cumulative distribution function of the standard normal distribution. 

For large $s_c/\sigma_S(l)$, the exceedance probability $\bar{\Phi}$ is exponentially small, and $N_{\mathrm{ind}} \bar{\Phi} \ll 1$. 

So, we can Taylor expand to first order to get the approximation
\begin{equation}
	\mathbb{P}\left( S_{\max}(l) > s_c \right) \approx 
	\frac{l}{\lambda_{\min}} \, \bar{\Phi}\!\left( 
	\frac{s_c}{\sigma_S(l)} \right),
	\label{eq:exceedanceProb}
\end{equation}
which is accurate in the regime relevant to earthquakes.

Then the rupture, having grown to size $L$, has survived the risk of 
arrest at all smaller scales. 

The expected number $\Lambda$ of potentially arresting bumps at scale $l$ is the number of independent windows of size $l$ along the total fault length $L_{\max}$, multiplied by the probability that a given window contains an arresting slope, which writes
\begin{equation}
	\Lambda(l) = \frac{L_{\max}}{l} \, \mathbb{P}\!\left( S_{\max}(l) 
	> s_c \right).
	\label{eq:hazardRate2}
\end{equation}
The cascade is then a survival process in scale space.
The rupture undergoes upscaling with a probability to reach size $L$ without encountering an arresting bump at any smaller scale given by the product of the survival probabilities ($1 -$ hazard rate $ = 1 - \Lambda(l) dl/l$) over all scales from $l_{\min}$ to $L$.

Assuming fast decorrelation from one scale to another, the rupture will upscale following a Poissonian process, which, in the continuum limit, gives survival probability given by 
\begin{equation}
	P_{\mathrm{surv}}(L) = \exp\!\left[ - \int_{l_{\min}}^{L} 
	\Lambda(l) \, \frac{dl}{l} \right].
	\label{eq:survivalIntegral2}
\end{equation}
Substituting the expressions for $\Lambda(l)$ and 
$\mathbb{P}(S_{\max} > s_c)$, we get
\begin{equation}
	\ln P_{\mathrm{surv}}(L) \approx - \frac{L_{\max}}{\lambda_{\min}} 
	\int_{l_{\min}}^{L} \frac{1}{l} \, \bar{\Phi}\!\left( 
	\frac{s_c}{C l^{-1/3}} \right) \, dl.
	\label{eq:logSurvival}
\end{equation}
To evaluate this integral, we introduce the dimensionless variable
$u(l) = s_c / \sigma_S(l) = (s_c / C) \, l^{1/3}$, so that
$l = (C u / s_c)^3$ and $dl/l = 3\, du/u$.  
Then, the integration limits become $u_{\min} = (s_c/C) \, l_{\min}^{1/3}$ and $u(L) = (s_c/C) \, L^{1/3}$, giving
\begin{equation}
	\ln P_{\mathrm{surv}}(L) \approx - \frac{3 L_{\max}}{\lambda_{\min}} 
	\int_{u_{\min}}^{u(L)} \bar{\Phi}(u) \, \frac{du}{u}.
	\label{eq:survivalU}
\end{equation}
For large $u$, $\bar{\Phi}(u) \sim (2\pi)^{-1/2} u^{-1} e^{-u^2/2}$, so
the integrand decays super-exponentially and the integral is dominated by
the contribution near the upper limit $u(L)$.  
We can then apply the steepest-descent technique to estimate the leading-order behaviour
\begin{equation}
	P_{\mathrm{surv}}(L) \approx \exp\!\left[ - k \left( 
	\frac{L}{L_{\mathrm{crit}}} \right)^{\beta} \right],
	\label{eq:WeibullForm}
\end{equation}
with $\beta = 1 - \zeta = 1/3$, so that 
$L_{\mathrm{crit}} = \left( \sigma_n C / \Delta\tau \right)^3$, and $k$ a
dimensionless constant.  

\section{Derivation of the general formula for $D_c$ from fractal contact dynamics}

Consider the fault as a collection of discrete contacts whose sizes $l$
follow the power-law distribution $n(l) \propto l^{-4/3}$ derived in the
main text. Each contact breaks when the local accumulated slip reaches
$d_c(l) = \max(\delta_{\mathrm{frac}}, \frac{C}{2\mu} l^{\zeta})$,
giving a lifetime
\begin{equation}
	t_{\mathrm{life}}(l) = \frac{d_c(l)}{V}
	\label{eq:lifetime}
\end{equation}
at sliding velocity $V$. Let $N(t,l,t_c)$ be the number density of
contacts of size $l$ and age $t_c$ (time elapsed since formation) at
time $t$. The age dynamics is governed by the McKendrick--von Foerster
equation
\begin{equation}
	\frac{\partial N}{\partial t} + \frac{\partial N}{\partial t_c} = 
	- \frac{N}{t_{\mathrm{life}}(l)} + \delta(t_c) \int_0^\infty 
	\frac{N(t, l, t_c')}{t_{\mathrm{life}}(l)} \, dt_c',
	\label{eq:McKendrick}
\end{equation}
where the left-hand side is the convective derivative along the age axis,
the first term on the right represents destruction of contacts at rate
$1/t_{\mathrm{life}}$, and the second term represents birth of new
contacts at age zero.

In the steady-state regime, the slip occurs at constant velocity $V$ with $\partial N / \partial t = 0$.
For $t_c > 0$ the delta function vanishes, and the equation reduces to
$dN_0/dt_c = -N_0/t_{\mathrm{life}}$, giving the exponential age
distribution $N_0(l,t_c) = N_0(l,0) \, e^{-t_c/t_{\mathrm{life}}(l)}$.
The constant $N_0(l,0)$ is determined by equating the birth rate to the
total death rate. 

Let $N_{\mathrm{tot}}(l) = \int_0^\infty N_0(l,t_c) dt_c$
be the total number density of contacts of size $l$. 
The total death rate is $\int_0^\infty (N_0(l,t_c)/t_{\mathrm{life}}) dt_c = N_{\mathrm{tot}}/t_{\mathrm{life}}$, and the birth rate is $N_0(l,0)$. 

Equating the two gives $N_0(l,0) = N_{\mathrm{tot}}(l)/t_{\mathrm{life}}(l)$, so that
\begin{equation}
	N_0(l,t_c) = \frac{N_{\mathrm{tot}}(l)}{t_{\mathrm{life}}(l)} 
	\, e^{-t_c / t_{\mathrm{life}}(l)}.
	\label{eq:ageDist}
\end{equation}
The total number of contacts of size $l$ is proportional to the geometric
size distribution: $N_{\mathrm{tot}}(l) = \rho \, n(l) \propto l^{-4/3}$ per unit area.

The shear strength of a single contact is assumed to grow logarithmically
with its age, as routinely observed in laboratory experiments:
\begin{equation}
	\tau_{\mathrm{contact}}(l,t_c) = \tau_0(l) + \bar{a} \ln\!\left( 
	1 + \frac{t_c}{\epsilon} \right),
	\label{eq:contactStrength}
\end{equation}
where $\tau_0(l)$ is the instantaneous strength, $\bar{a}$ is the logarithmic strengthening rate (stress unit), and $\epsilon$ is a
microscopic time cutoff ($\epsilon \ll t_{\mathrm{life}}$). 

The total macroscopic shear stress is the spatial average of the forces
carried by all contacts.  Let $N_0(l,t_c)$ be the number of contacts
per unit fault area, per unit size $l$, and per unit age $t_c$.  Each
contact of size $l$ has an area $A_c(l) \propto l^2$, so the force it
sustains is $\tau_{\mathrm{contact}}(l,t_c) \, A_c(l)$.  Summing over
all contacts and normalising by the fault area gives
\begin{equation}
	\tau = \int_{l_{\min}}^{L_{\max}} A_c(l) \int_0^\infty N_0(l,t_c) \,
	\tau_{\mathrm{contact}}(l,t_c) \, dt_c \, dl.
	\label{eq:totalTau}
\end{equation}

Inserting the steady‑state age distribution (\ref{eq:ageDist}) and the
logarithmic contact‑strengthening law (\ref{eq:contactStrength}) yields
\begin{equation}
	\tau = \int A_c(l) N_{\mathrm{tot}}(l)
	\Big[ \tau_0(l) + \bar{a} \, \mathcal{I}(l) \Big] dl,
	\label{eq:tauSplit}
\end{equation}
where $N_{\mathrm{tot}}(l) = \int_0^\infty N_0(l,t_c) dt_c$ is the
number of contacts per unit fault area per unit size,
and
\begin{equation}
	\mathcal{I}(l) = \frac{1}{t_{\mathrm{life}}(l)} 
	\int_0^\infty e^{-t_c/t_{\mathrm{life}}} \ln\!\left(1 + \frac{t_c}{\epsilon}\right) dt_c.
	\label{eq:Idef}
\end{equation}

Evaluating the age integral with $x = t_c/t_{\mathrm{life}}$ and
$t_{\mathrm{life}} \gg \epsilon$ 
Evaluating the age integral with the substitution $x = t_c / t_{\mathrm{life}}$ and using
$t_{\mathrm{life}} \gg \epsilon$, we obtain
\begin{equation}
	\begin{aligned}
		\mathcal{I}(l) &\approx \int_0^\infty e^{-x} \Big[ \ln\!\left(\frac{t_{\mathrm{life}}}{\epsilon}\right) + \ln x \Big] dx \\
		&= \ln\!\left(\frac{t_{\mathrm{life}}}{\epsilon}\right) \int_0^\infty e^{-x} dx + \int_0^\infty e^{-x} \ln x \, dx \\
		&= \ln\!\left(\frac{d_c(l)}{V \epsilon}\right) - \gamma,
	\end{aligned}
	\label{eq:Ieval}
\end{equation}
where $\gamma \approx 0.577$ is the Euler-Mascheroni constant, and we used
$t_{\mathrm{life}} = d_c / V$.

Substituting back and collecting terms:
\begin{equation}
	\begin{aligned}
		\tau &= \int A_c(l) N_{\mathrm{tot}}(l) \tau_0(l) dl \\
		&\quad + \bar{a} \int A_c(l) N_{\mathrm{tot}}(l) \ln d_c(l) dl \\
		&\quad - \bar{a} \ln V \int A_c(l) N_{\mathrm{tot}}(l) dl
		+ \text{constant}.
	\end{aligned}
	\label{eq:tauExpanded}
\end{equation}

If we define the total contact area per unit fault area, $\rho_A = \int A_c(l) N_{\mathrm{tot}}(l) dl$ is a dimensionless quantity, while the third term is $-\bar{a} \rho_A \ln V$, the direct velocity effect.
Comparing with the standard form $a \sigma_n \ln(V/V_0)$ in the rate-and-state friction laws gives
\begin{equation}
	a = \bar{a} \rho_A / \sigma_n.  
\end{equation}

The second term is the state contribution. Comparing with the rate-and-state steady-state stress
$\tau_{ss} = \tau_* + a \sigma_n \ln(V/V_*) + b \sigma_n \ln(V_* \theta_{ss} / D_c)$, where $\theta_{ss} = D_c/V$, we identify
\begin{equation}
	\begin{aligned}
		b \sigma_n \ln (D_c/d_0) &\propto \bar{a} \int A_c(l) N_{\mathrm{tot}}(l) \ln (d_c(l)/d_0) dl \\
		&= \bar{a} \rho_A \frac{\int A_c(l) N_{\mathrm{tot}}(l) \ln (d_c(l)/d_0) dl}
		{\int A_c(l) N_{\mathrm{tot}}(l) dl}.
	\end{aligned}
	\label{eq:stateIdentification}
\end{equation}

Using $\bar{a} \rho_A = a \sigma_n$, the normal stress cancels, providing the final formula for the rate-and-state critical slip distance in terms of the underlying structural statistical properties of faults
\begin{equation}
	D_c = d_0\exp\!\left[ \frac{a}{b}\,
		\frac{\int_{l_{\min}}^{L_{\max}} A_c(l) N_{\mathrm{tot}}(l) \ln (d_c(l)/d_0) dl}
		{\int_{l_{\min}}^{L_{\max}} A_c(l) N_{\mathrm{tot}}(l) dl} \right], 
	\label{eq:DcGeneral}
\end{equation}
where $d_0$ arises from logarithm normalization. 
If the reference length is set implicit in the units, we get 
\begin{equation}
	D_c = \exp\!\left[ \frac{a}{b}\,
	\frac{\int_{l_{\min}}^{L_{\max}} A_c(l) N_{\mathrm{tot}}(l) \ln d_c(l) dl}
	{\int_{l_{\min}}^{L_{\max}} A_c(l) N_{\mathrm{tot}}(l) dl} \right].
	\label{eq:DcGeneral2}
\end{equation}
Therefore, $D_c$ represents the geometric mean of the size-dependent critical slip distances $d_c(l)$, weighted by the contact area distribution $A_c(l) N_{\mathrm{tot}}(l)$.
Indeed, its formula can be written explicitly in a more transparent posing $a/b \approx 1$ (as it is in most scenarios) as 
\begin{equation}
	D_c = \left( \prod_{l} d_c(l)^{\,W(l)} \right)^{\dfrac{1}{\int W(l)\,dl}},~~W(l) = A_c(l) N_{\mathrm{tot}}(l),
	\label{eq:DcGeometricMean}
\end{equation}
where the product is performed over all contact sizes and the weight
$W(l)$ depends on the contact area distribution.                    
Written in this way, it is clear that $D_c$ is not the critical slip of any single contact but the average over the entire fractal population, with larger contacts contributing more heavily because of their greater area.

The detailed physical interpretation of this formula is discussed in the main text.

\section{Derivation of the aging law}
When the sliding velocity changes, the contact population relaxes toward a new steady state. 

In a quasi-static approximation, where velocity changes are slow compared to the lifetime of individual contacts, the age distribution keeps its exponential form
\begin{equation}
  N(t_c, t) = [N_{\mathrm{tot}} / t_{\mathrm{life}}(t)] \, e^{-t_c / t_{\mathrm{life}}(t)}
\end{equation}
but now with the time-dependent mean lifetime 
$t_{\mathrm{life}}(t) = D_c / V(t)$. 
Indeed, the mean contact age
$\langle t_c \rangle = \frac{1}{N_{\mathrm{tot}}} \int_0^\infty t_c N(t_c, t) dt_c$
evolves according to two competing effects: contacts age at a unit rate, while the death and birth process continually replaces older contacts with newborn ones of age zero, pulling the mean age downward. The McKendrick-von Foerster equation gives the exact balance between these two effects as
\begin{equation}
	\frac{d\langle t_c \rangle}{dt} = 1 - \frac{\langle t_c \rangle}{t_{\mathrm{life}}(t)},
	\label{eq:ageEvolution}
\end{equation}
where the second term means that a fraction $1/t_{\mathrm{life}}$ of
the population is replaced per unit time, and older contacts are
preferentially removed because their longer age places them nearer to the
end of their lifetime.

In steady state, the two terms balance and
$\langle t_c \rangle = t_{\mathrm{life}} = D_c/V$.  

If we identify the rate-and-state variable as $\theta = \langle t_c \rangle$, and using $t_{\mathrm{life}} = D_c/V$, Eq.~(\ref{eq:ageEvolution}) becomes
\begin{equation}
	\frac{d\theta}{dt} = 1 - \frac{V\theta}{D_c},
	\label{eq:agingLawDerived}
\end{equation}
which is exactly the Dieterich aging law. The critical slip distance
$D_c$ appearing in this equation is the same structural average derived in
Eq.~(\ref{eq:DcGeneral2}), linking the macroscopic state evolution to the
underlying fault roughness.

\begin{thebibliography}{99}
	\onecolumngrid
	\section*{References}
	\twocolumngrid
\bibitem{Cocco2002}
M.~Cocco and A.~Bizzarri,
On the slip-weakening behavior of rate- and state-dependent constitutive laws,
\emph{Geophys. Res. Lett.} \textbf{29}, 1516 (2002).

\bibitem{Tinti2005}
E.~Tinti, E.~Fukuyama, A.~Piatanesi, and M.~Cocco,
A kinematic source-time function compatible with earthquake dynamics,
\emph{Bull. Seismol. Soc. Am.} \textbf{95}, 1211--1223 (2005).

\bibitem{Dieterich1979}
J.~H.~Dieterich,
Modeling of rock friction: 1. Experimental results and constitutive equations,
\emph{J. Geophys. Res.} \textbf{84}, 2161--2168 (1979).

\bibitem{Marone1998}
C.~Marone,
Laboratory-derived friction laws and their application to seismic faulting,
\emph{Annu. Rev. Earth Planet. Sci.} \textbf{26}, 643--696 (1998).

\bibitem{Papageorgiou1983}
A.~S.~Papageorgiou and K.~Aki,
A specific barrier model for the quantitative description of inhomogeneous faulting and the prediction of strong ground motion. I. Description of the model,
\emph{Bull. Seismol. Soc. Am.} \textbf{73}, 693--722 (1983).

\bibitem{Ellsworth2003}
W.~L.~Ellsworth and G.~C.~Beroza,
Seismic evidence for an earthquake nucleation phase,
\emph{Science} \textbf{268}, 851--855 (1995).

\bibitem{Scholz1988}
C.~H.~Scholz,
The critical slip distance for seismic faulting,
\emph{Nature} \textbf{336}, 761--763 (1988).

\bibitem{Ide1997}
S.~Ide and M.~Takeo,
Determination of constitutive relations of fault slip based on seismic wave analysis,
\emph{J. Geophys. Res.} \textbf{102}, 27379--27391 (1997).

\bibitem{Galetzka2015}
J.~Galetzka, D.~Melgar, J.~F.~Genrich, J.~Geng, S.~Owen, E.~O.~Lindsey, X.~Xu, Y.~Bock, J.-P.~Avouac, L.~B.~Adhikari, B.~N.~Upreti, B.~Pratt-Sitaula, T.~N.~Bhattarai, B.~P.~Sitaula, A.~Moore, K.~W.~Hudnut, W.~Szeliga, J.~Normandeau, M.~Fend, M.~Flouzat, L.~Bollinger, P.~Shrestha, B.~Koirala, U.~Gautam, M.~Bhatterai, R.~Gupta, T.~Kandel, C.~Timsina, S.~N.~Sapkota, S.~Rajaure, and N.~Maharjan,
Slip pulse and resonance of the Kathmandu basin during the 2015 Gorkha earthquake, Nepal,
\emph{Science} \textbf{349}, 1091--1095 (2015).

\bibitem{Zaccagnino2026}
D.~Zaccagnino,
Scale-dependent earthquake nucleation: Implications for seismicity, tectonics and laboratory experiments,
\emph{Tectonophysics} 231156 (2026).

\bibitem{Zaccagnino2025}
D.~Zaccagnino, O.~Bruno, and C.~Doglioni,
Spatial scale dependence of fault physical parameters and its implications for the analysis of earthquake dynamics from the lab to fault systems,
\emph{Earth Planet. Sci. Lett.} \textbf{666}, 119481 (2025).

\bibitem{DalZilio2023}
L.~Dal~Zilio, P.~A.~Selvadurai, J.~P.~Ampuero, E.~Tinti, M.~Cocco, F.~Cappa, and B.~Team,
Can earthquakes nucleate on nominally stable velocity-strengthening faults?,
\emph{EGU General Assembly Conference Abstracts}, EGU-11960 (2023).

\bibitem{Barbery2025}
M.~Barbery, G.~Hirth, and T.~Tullis,
Strong asperities nucleate earthquakes on laboratory faults,
\emph{Geology} \textbf{53}, 420--424 (2025).

\bibitem{Li2025}
M.~Li, A.~R.~Niemeijer, and Y.~van Dinther,
Frictional healing and induced earthquakes on conventionally stable faults,
\emph{Nat. Commun.} \textbf{16}, 9140 (2025).

\bibitem{Ross2020}
Z.~E.~Ross, E.~S.~Cochran, D.~T.~Trugman, and J.~D.~Smith,
3D fault architecture controls the dynamism of earthquake swarms,
\emph{Science} \textbf{368}, 1357--1361 (2020).

\bibitem{Cochran2023}
E.~S.~Cochran, M.~T.~Page, N.~J.~Van~Der~Elst, Z.~E.~Ross, and D.~T.~Trugman,
Fault roughness at seismogenic depths and links to earthquake behavior,
\emph{The Seismic Record} \textbf{3}, 37--47 (2023).

\bibitem{Lee2024}
J.~Lee, V.~C.~Tsai, G.~Hirth, A.~Chatterjee, and D.~T.~Trugman,
Fault-network geometry influences earthquake frictional behaviour,
\emph{Nature} \textbf{631}, 106--110 (2024).

\bibitem{Harbord2017}
C.~W.~Harbord, S.~B.~Nielsen, N.~De~Paola, and R.~E.~Holdsworth,
Earthquake nucleation on rough faults,
\emph{Geology} \textbf{45}, 931 (2017).

\bibitem{Dong2024}
P.~Dong, Z.~Wang, Y.~Xu, and K.~Xia,
Effects of fault roughness on estimating critical slip-weakening distance from fault slip history: A laboratory study,
\emph{Tectonophysics} \textbf{885}, 230419 (2024).

\bibitem{Ruina1983}
A.~L.~Ruina,
Slip instability and state variable friction laws,
\emph{J. Geophys. Res.} \textbf{88}, 10359--10370 (1983).

\bibitem{Rice1983}
J.~R.~Rice and A.~L.~Ruina,
Stability of steady frictional slipping,
\emph{J. Appl. Mech.} \textbf{50}, 343--349 (1983).

\bibitem{DeRubeis1996}
V.~De~Rubeis, R.~Hallgass, V.~Loreto, G.~Paladin, L.~Pietronero, and P.~Tosi,
Self-affine asperity model for earthquakes,
\emph{Phys. Rev. Lett.} \textbf{76}, 2599 (1996).

\bibitem{Milanese2019}
E.~Milanese, T.~Brink, R.~Aghababaei, and J.~F.~Molinari,
Emergence of self-affine surfaces during adhesive wear,
\emph{Nat. Commun.} \textbf{10}, 1116 (2019).

\bibitem{Schmittbuhl1993}
J.~Schmittbuhl, S.~Gentier, and S.~Roux,
Field measurements of the roughness of fault surfaces,
\emph{Geophys. Res. Lett.} \textbf{20}, 639--641 (1993).

\bibitem{Pozzi2022}
G.~Pozzi, M.~M.~Scuderi, E.~Tinti, M.~Nazzari, and C.~Collettini,
The role of fault rock fabric in the dynamics of laboratory faults,
\emph{J. Geophys. Res.} \textbf{127}, e2021JB023779 (2022).

\bibitem{Fang2013}
Z.~Fang and E.~M.~Dunham,
Additional shear resistance from fault roughness and stress levels on geometrically complex faults,
\emph{J. Geophys. Res.} \textbf{118}, 3642--3654 (2013).

\bibitem{Weiss2016}
J.~Weiss, V.~Pellissier, D.~Marsan, L.~Arnaud, and F.~Renard,
Cohesion versus friction in controlling the long-term strength of a self-healing experimental fault,
\emph{J. Geophys. Res.} \textbf{121}, 8523--8547 (2016).

\bibitem{Gabrielov1996}
A.~Gabrielov, V.~Keilis-Borok, and D.~D.~Jackson,
Geometric incompatibility in a fault system,
\emph{Proc. Natl. Acad. Sci.} \textbf{93}, 3838--3842 (1996).

\bibitem{Nielsen1998}
S.~B.~Nielsen and L.~Knopoff,
The equivalent strength of geometrical barriers to earthquakes,
\emph{J. Geophys. Res.} \textbf{103}, 9953--9965 (1998).

\bibitem{Power1987}
W.~L.~Power, T.~E.~Tullis, S.~R.~Brown, G.~N.~Boitnott, and C.~H.~Scholz,
Roughness of natural fault surfaces,
\emph{Geophys. Res. Lett.} \textbf{14}, 29--32 (1987).

\bibitem{Schmittbuhl1995}
J.~Schmittbuhl, F.~Renard, J.~P.~Gratier, and R.~Toussaint,
Roughness of stylolites: Implications of 3D high resolution topography measurements,
\emph{Phys. Rev. Lett.} \textbf{93}, 238501 (2004).

\bibitem{Sagy2007}
A.~Sagy, E.~E.~Brodsky, and G.~J.~Axen,
Evolution of fault-surface roughness with slip,
\emph{Geology} \textbf{35}, 283--286 (2007).

\bibitem{Candela2009}
T.~Candela, F.~Renard, Y.~Klinger, K.~Mair, J.~Schmittbuhl, and E.~E.~Brodsky,
Roughness of fault surfaces over nine decades of length scales,
\emph{J. Geophys. Res.} \textbf{117}, B08409 (2012).

\bibitem{Brodsky2011}
E.~E.~Brodsky, J.~J.~Gilchrist, A.~Sagy, and C.~Collettini,
Faults smooth gradually as a function of slip,
\emph{Earth Planet. Sci. Lett.} \textbf{302}, 185--193 (2011).

\bibitem{KPZ1986}
M.~Kardar, G.~Parisi, and Y.-C.~Zhang,
Dynamic scaling of growing interfaces,
\emph{Phys. Rev. Lett.} \textbf{56}, 889--892 (1986).

\bibitem{Barabasi1995}
A.-L.~Barab\'asi and H.~E.~Stanley,
\emph{Fractal Concepts in Surface Growth},
Cambridge University Press, Cambridge, 1995.

\bibitem{Tullis1986}
T.~E.~Tullis and J.~D.~Weeks,
Constitutive behavior and stability of frictional sliding of granite,
\emph{Pure Appl. Geophys.} \textbf{124}, 383--414 (1986).

\bibitem{Kilgore1993}
B.~D.~Kilgore, M.~L.~Blanpied, and J.~H.~Dieterich,
Velocity dependent friction of granite over a wide range of conditions,
\emph{Geophys. Res. Lett.} \textbf{20}, 903--906 (1993).

\bibitem{Gibowicz1994}
S.~J.~Gibowicz and A.~Kijko,
\emph{An Introduction to Mining Seismology},
Academic Press, 1994.

\bibitem{Mair1999}
K.~Mair and C.~Marone,
Friction of simulated fault gouge for a wide range of velocities and normal stresses,
\emph{J. Geophys. Res.} \textbf{104}, 28899--28914 (1999).

\bibitem{Ohnaka1999}
M.~Ohnaka and L.~F.~Shen,
Scaling of the shear rupture process from nucleation to dynamic propagation: Implications of geometric irregularity of the rupturing surfaces,
\emph{J. Geophys. Res.} \textbf{104}, 817--844 (1999).

\bibitem{Ohnaka2000}
M.~Ohnaka,
A physical scaling relation between the size of an earthquake and its nucleation zone size,
\emph{Pure Appl. Geophys.} \textbf{157}, 2259--2282 (2000).

\bibitem{Richardson2002}
E.~Richardson and T.~H.~Jordan,
Seismicity in deep gold mines of South Africa: Implications for tectonic earthquakes,
\emph{Bull. Seismol. Soc. Am.} \textbf{92}, 1766--1782 (2002).

\bibitem{Mikumo2003}
T.~Mikumo, K.~B.~Olsen, E.~Fukuyama, and Y.~Yagi,
Stress-breakdown time and slip-weakening distance inferred from slip-velocity functions on earthquake faults,
\emph{Bull. Seismol. Soc. Am.} \textbf{93}, 264--282 (2003).

\bibitem{Chambon2006}
G.~Chambon, J.~Schmittbuhl, and A.~Corfdir,
Frictional response of a thick gouge sample: 1. Mechanical measurements and microstructures,
\emph{J. Geophys. Res.} \textbf{111}, B09308 (2006).

\bibitem{Leeman2016}
J.~R.~Leeman, D.~M.~Saffer, M.~M.~Scuderi, and C.~Marone,
Laboratory observations of slow earthquakes and the spectrum of tectonic fault slip modes,
\emph{Nat. Commun.} \textbf{7}, 11104 (2016).

\bibitem{Scuderi2017}
M.~M.~Scuderi and C.~Collettini,
The role of fluid pressure in induced vs.\ triggered seismicity: Insights from rock deformation experiments on carbonates,
\emph{Sci. Rep.} \textbf{6}, 24852 (2016).

\bibitem{Rubino2017}
V.~Rubino, A.~J.~Rosakis, and N.~Lapusta,
Understanding dynamic friction through spontaneously evolving laboratory earthquakes,
\emph{Nat. Commun.} \textbf{8}, 15991 (2017).

\bibitem{Ando2018}
Y.~Kaneko, E.~Fukuyama, and I.~J.~Hamling,
Slip-weakening distance and energy budget inferred from near-fault ground deformation during the 2016 $M_w$~7.8 Kaik\={o}ura earthquake,
\emph{Geophys. Res. Lett.} \textbf{44}, 4765--4773 (2017).

\bibitem{Ikari2019}
M.~J.~Ikari,
Laboratory slow slip events in natural geologic materials,
\emph{Geophys. J. Int.} \textbf{218}, 354--387 (2019).

\bibitem{McLaskey2019}
G.~C.~McLaskey,
Earthquake initiation from laboratory observations and implications for foreshocks,
\emph{J. Geophys. Res.} \textbf{124}, 12882--12904 (2019).

\bibitem{Latour2025}
S.~Latour, M.~Lebihain, H.~S.~Bhat, C.~Twardzik, Q.~Bletery, K.~W.~Hudnut, and F.~Passel\`e{ }gue,
Direct estimation of earthquake source properties from a single CCTV camera,
\emph{Science} \textbf{390}, 463--467 (2025).

\bibitem{Lawn1993}
B.~R.~Lawn,
\emph{Fracture of Brittle Solids}, 2nd ed.,
Cambridge University Press, Cambridge, 1993.

\bibitem{Heaton1990}
T.~H.~Heaton,
Evidence for and implications of self-healing pulses of slip in earthquake rupture,
\emph{Phys. Earth Planet. Inter.} \textbf{64}, 1--20 (1990).

\bibitem{Zheng1998}
G.~Zheng and J.~R.~Rice,
Conditions under which velocity-weakening friction allows a self-healing versus a cracklike mode of rupture,
\emph{Bull. Seismol. Soc. Am.} \textbf{88}, 1466--1483 (1998).

\bibitem{Ampuero2008}
J.-P.~Ampuero and Y.~Ben-Zion,
Cracks, pulses and macroscopic asymmetry of dynamic rupture on a bimaterial interface with velocity-weakening friction,
\emph{Geophys. J. Int.} \textbf{173}, 674--692 (2008).

\bibitem{Gabriel2012}
A.-A.~Gabriel, J.-P.~Ampuero, L.~A.~Dalguer, and P.~M.~Mai,
The transition of dynamic rupture styles in elastic media under velocity-weakening friction,
\emph{J. Geophys. Res.} \textbf{117}, B09311 (2012).

\bibitem{Weng2017}
H.~Weng and H.~Yang,
Seismogenic width controls aspect ratios of earthquake ruptures,
\emph{Geophys. Res. Lett.} \textbf{44}, 2725--2732 (2017).

\bibitem{Weng2019}
H.~Weng and J.-P.~Ampuero,
The dynamics of elongated earthquake ruptures,
\emph{J. Geophys. Res.} \textbf{124}, 8584--8610 (2019).

\bibitem{Martel1997}
S.J.~Martel,
Effects of cohesive zones on small faults and implications for secondary fracturing and fault trace geometry,
\emph{J. Struct. Geol.}, \textbf{19}, 835--847 (1997).

\bibitem{Hsu1969}
K.J.~Hsu,
Role of cohesive strength in the mechanics of overthrust faulting and of landsliding,
\emph{Geol. Soc. Am. Bull.}, \textbf{80}, 927--952 (1969).

\bibitem{Tenthorey2006}
E.~Tenthorey and S.F.~Cox,
Cohesive strengthening of fault zones during the interseismic period: An experimental study,
\emph{J. Geophys. Res.: Solid Earth}, \textbf{111}, B09202 (2006).

\bibitem{Hirata1989}
T.~Hirata,
Fractal dimension of fault systems in Japan: fractal structure in rock fracture geometry at various scales,
\emph{Pure Appl. Geophys.}, \textbf{131}, 157--170 (1989).

\bibitem{Kagan1991}
Y.Y.~Kagan,
Fractal dimension of brittle fracture,
\emph{J. Nonlinear Sci.}, \textbf{1}, 1--16 (1991).

\bibitem{Cowie1995}
P.A.~Cowie, D.~Sornette, and C.~Vanneste,
Multifractal scaling properties of a growing fault population,
\emph{Geophys. J. Int.}, \textbf{122}, 457--469 (1995).

\bibitem{BenZion2003}
Y.~Ben-Zion and C.G.~Sammis,
Characterization of fault zones,
\emph{Pure Appl. Geophys.}, \textbf{160}, 677--715 (2003).

\bibitem{Mckendrick1925}
A.G.~M'Kendrick,
Applications of mathematics to medical problems,
\emph{Proc. Edinburgh Math. Soc.}, \textbf{44}, 98--130 (1925).

\bibitem{VonFoerster1959}
H.~von~Foerster,
Some remarks on changing populations,
\emph{The Kinetics of Cellular Proliferation}, 382--407 (1959).

\bibitem{VenegasAravena2024}
P.~Venegas-Aravena, J.~G.~Crempien, and R.~J.~Archuleta,
Fractal spatial distributions of initial shear stress and frictional properties on faults and their impact on dynamic earthquake rupture,
\emph{Bull. Seismol. Soc. Am.} \textbf{114}, 1444--1465 (2024).

\bibitem{VenegasAravena2025}
P.~Venegas-Aravena and D.~Zaccagnino,
Large earthquakes are more predictable than smaller ones,
\emph{Seismica} \textbf{4} (2025).

\bibitem{Caniven2017}
Y.~Caniven, S.~Dominguez, R.~Soliva, M.~Peyret, R.~Cattin, and F.~Maerten,
Relationships between along-fault heterogeneous normal stress and fault slip patterns during the seismic cycle: Insights from a strike-slip fault laboratory model,
\emph{Earth Planet. Sci. Lett.} \textbf{480}, 147--157 (2017).

\bibitem{Dolan2017}
J.~F.~Dolan and B.~J.~Meade,
A comparison of geodetic and geologic rates prior to large strike-slip earthquakes: A diversity of earthquake-cycle behaviors?,
\emph{Geochem. Geophys. Geosyst.} \textbf{18}, 4426--4436 (2017).

\bibitem{Zaccagnino2021}
D.~Zaccagnino, L.~Telesca, and C.~Doglioni,
Different fault response to stress during the seismic cycle,
\emph{Appl. Sci.} \textbf{11}, 9596 (2021).

\bibitem{Barbot2023}
S.~Barbot,
Constitutive behavior of rocks during the seismic cycle,
\emph{AGU Advances} \textbf{4}, e2023AV000972 (2023).

\bibitem{Zaccagnino2026b}
D.~Zaccagnino,
Short-sighted faults: A new model for earthquake recurrence and crustal memory,
\emph{Terra Nova} \textbf{38}, 241--250 (2026).

\bibitem{Brodsky2026}
E.~E.~Brodsky and G.~Farge,
How earthquakes organize stress,
\emph{Proc. Natl. Acad. Sci. U.S.A.} \textbf{123}, e2530754123 (2026).

\bibitem{Melgar2017}
D.~Melgar and G.~P.~Hayes,
Systematic observations of the slip pulse properties of large earthquake ruptures,
\emph{Geophys. Res. Lett.} \textbf{44}, 9691--9698 (2017).

\bibitem{Lambert2021}
V.~Lambert, N.~Lapusta, and S.~Perry,
Propagation of large earthquakes as self-healing pulses or mild cracks,
\emph{Nature} \textbf{591}, 252--258 (2021).

\bibitem{Brodsky2001}
E.~E.~Brodsky and H.~Kanamori,
Elastohydrodynamic lubrication of faults,
\emph{J. Geophys. Res.} \textbf{106}, 16357--16374 (2001).

\bibitem{DePaola2011}
N.~De~Paola, T.~Hirose, T.~Mitchell, G.~Di~Toro, C.~Viti, and T.~Shimamoto,
Fault lubrication and earthquake propagation in thermally unstable rocks,
\emph{Geology} \textbf{39}, 35--38 (2011).

\bibitem{Pozzi2021}
G.~Pozzi, N.~De~Paola, S.~B.~Nielsen, R.~E.~Holdsworth, T.~Tesei, M.~Thieme, and S.~Demouchy,
Coseismic fault lubrication by viscous deformation,
\emph{Nat. Geosci.} \textbf{14}, 437--442 (2021).

\bibitem{Gnedenko1968}
B.~V.~Gnedenko,
\emph{The Theory of Probability},
translated by B.~D.~Seckler,
CUP Archive, 1968.
	
\end{thebibliography}
\end{document}